\documentclass{article}

\usepackage{microtype}
\usepackage{graphicx}
\usepackage{subfigure}
\usepackage{booktabs} 
\usepackage{amsmath}
\usepackage{amssymb}
\usepackage{listings}
\usepackage{tcolorbox}
\usepackage{siunitx}
\usepackage{makecell}
\usepackage{multirow}
\usepackage{xcolor}
\usepackage{tikz}                   
\usepackage{pgfplots}              
\usepackage{pgfplotstable}         
\usepackage{subcaption}

\pgfplotsset{compat=1.18}       
\usepgfplotslibrary{colormaps}     
\usetikzlibrary{plotmarks}         
\usepgfplotslibrary{fillbetween}  

\lstdefinelanguage{PyCustom}{
  language=Python,
  morekeywords={self,AssemblingMachine,Prototype,Direction,Position,Entity,TypeError},
  sensitive=true,
  morecomment=[l]{\#},
  morestring=[b]",
  morestring=[b]',
}

\definecolor{codebackground}{rgb}{0.98, 0.98, 0.98}
\definecolor{codegray}{rgb}{0.5,0.5,0.5}
\definecolor{darkblue}{rgb}{0.1,0.1,0.9}
\definecolor{darkred}{rgb}{0.6,0.0,0.0}

\usepackage{hyperref}

\usepackage{icml2024} 

\usepackage{amsmath}
\usepackage{amssymb}
\usepackage{mathtools}
\usepackage{amsthm}

\usepackage[capitalize,noabbrev]{cleveref}

\theoremstyle{plain}

\theoremstyle{definition}

\theoremstyle{remark}

\def\hyph{-\penalty0\hskip0pt\relax}

\usepackage{pifont}
\usepackage{array}
\usepackage{booktabs}

\definecolor{success}{RGB}{0,150,0}    
\definecolor{failure}{RGB}{200,0,0}    
\newcommand{\cmark}{\textcolor{success}{\ding{51}}\kern-3pt}
\newcommand{\xmark}{\textcolor{failure}{\ding{55}}\kern-3pt}
\newcommand{\tickgroup}[2]{#1\kern4pt(#2)}

\makeatletter
\gdef\isaccepted{1}
\renewcommand{\Notice@String}{Preliminary work. Under review.}
\makeatother

\icmltitlerunning{Factorio Learning Environment}

\begin{document}

\twocolumn[
\icmltitle{Factorio Learning Environment}

\icmlsetsymbol{equal}{*}
\begin{icmlauthorlist}
\icmlauthor{Jack Hopkins*}{yyy}
\icmlauthor{Mart Bakler*}{yyy}
\icmlauthor{Akbir Khan}{ant}
\end{icmlauthorlist}

\icmlaffiliation{yyy}{Independent Researcher}
\icmlaffiliation{ant}{Anthropic}

\icmlcorrespondingauthor{Jack Hopkins}{jack.hopkins@me.com}

\icmlkeywords{Machine Learning, ICML}

\vskip 0.3in
]



\printAffiliationsAndNotice{\icmlEqualContribution} 

\begin{abstract}
Large Language Models (LLMs) are rapidly saturating existing benchmarks, necessitating new open-ended evaluations. We introduce the Factorio Learning Environment (FLE), based on the game of Factorio, that tests agents in long-term planning, program synthesis, and resource optimization. FLE provides exponentially scaling challenges---from basic automation to complex factories processing millions of resource units per second. We provide two settings: (1) \textit{lab-play} consisting of 24 structured tasks with fixed resources, and (2) \textit{open-play} with the unbounded task of building the largest factory on an procedurally generated map. We demonstrate across both settings that models still lack strong spatial reasoning. In lab-play, we find that LLMs exhibit promising short-horizon skills, yet are unable to operate effectively in constrained environments, reflecting limitations in error analysis. In open-play, while LLMs discover automation strategies that improve growth (e.g electric-powered drilling), they fail to achieve complex automation (e.g electronic-circuit manufacturing). We release FLE as an open-source platform\footnote{
https://github.com/JackHopkins/factorio-learning-environment}.
\end{abstract}

\section{Introduction}
\label{submission}

Large Language Models (LLMs) have demonstrated remarkable capabilities at solving complex question-answer (QA) problems, saturating benchmarks in factual recollection \citep{hendrycks2021measuring}, reasoning \citep{cobbe2021training} and code prediction \citep{chen2021evaluating}.

The strong performance across these diverse tasks suggests that LLMs have developed sophisticated reasoning capabilities, leading researchers to explore whether models can act as autonomous agents \citep{yang2023auto}. This has motivated a number of new agentic benchmarks focusing on long-term planning \citep{liu2023agentbench, ruan2023tptu}, learning in complex environments \citep{paglieri2024balrog, jimenez2023swe} and reliably learning from mistakes \citep{xing2024understanding,yamada2023evaluating,kambhampati2024llms}. However, similar to QA settings, these agentic benchmarks are likely to face saturation due to their natural completion states; which impose an upper bound on performance and limit our ability to differentiate superhuman models.

\begin{figure}
\centering
\includegraphics[width=0.45\textwidth]{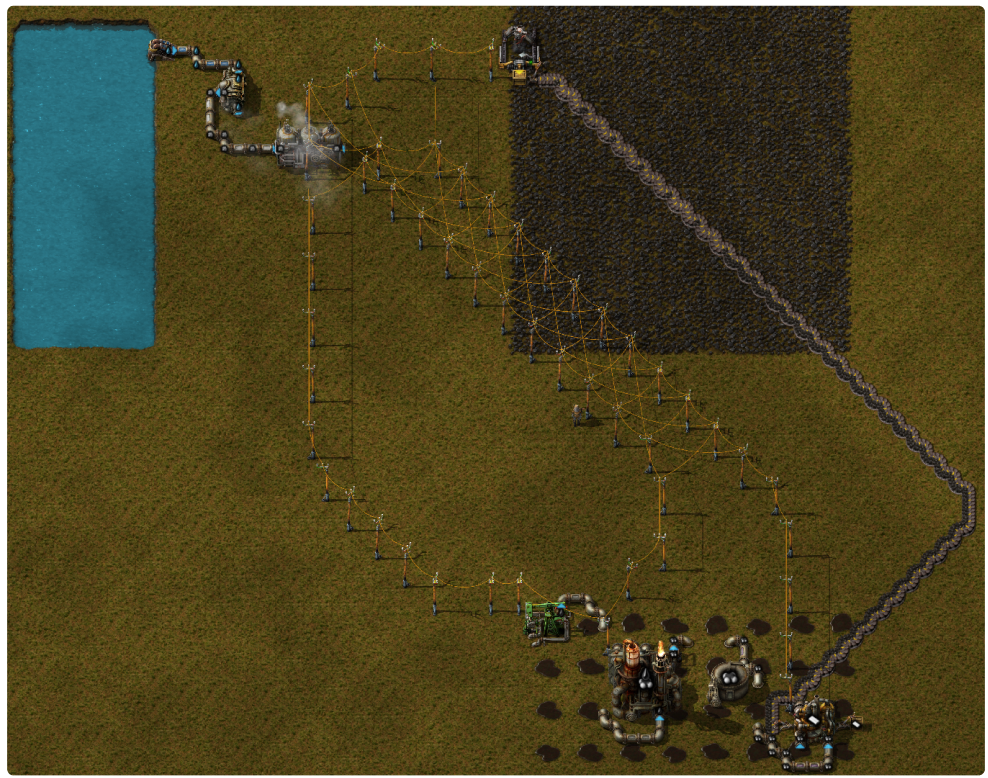}
\caption{\textbf{A plastic bar factory created by Claude 3.5 Sonnet in lab-play.} The factory consists of a electricity steam generator (top-left), a coal mine (top), a crude-oil to petroleum gas pipeline (bottom) and a chemical plant  (bottom-right). The chemical plant creates plastic bars using the coal and petroleum gas as inputs. By themselves, the cumulative raw resources generate a production score of $224$. With this specific layout, the factory creates $40$ plastic bars per $60$ in-game seconds, for a production score of $352$.}
\label{fig:example_factory}
\end{figure}

We introduce the \textbf{Factorio Learning Environment} (FLE): a novel evaluation framework built upon the game of Factorio that uniquely addresses this limitation by enabling unbounded agent evaluation with no natural completion state. In this environment, agents must navigate rapidly scaling challenges from basic resource gathering to complex automation while managing an exponentially scaling technology tree - creating natural curricula for evaluating increasingly capable agents. 

Agents are tasked with producing the largest possible factory, whose performance is measured through production throughput, which ranges from early-game rates of $\sim$30 resources/minute to advanced systems processing millions of resources/second. 
This enables us to meaningfully differentiate agents by measuring the order of magnitude of resources that they can produce, avoiding saturation by agents even as models become dramatically more capable.

Existing resource management environments such as Minecraft \citep{guss2019minerl} or Nethack \citep{kuttler2020nethack} do not demand the precise industrial optimization present in Factorio. For resource processing chains, producing basic electronic circuits (an early-game staple) requires coordinating 10+ machines processing approximately 15 items per minute. For example, a single rocket component requires orchestrating 60+ interlinked machines manufacturing 1000+ items per minute. The precision required, where a single misaligned machine can cause a factory-wide gridlock, creates a natural curriculum, testing both basic automation and advanced system optimization. 

\begin{figure*}[t]
\centering
\includegraphics[width=1\textwidth]{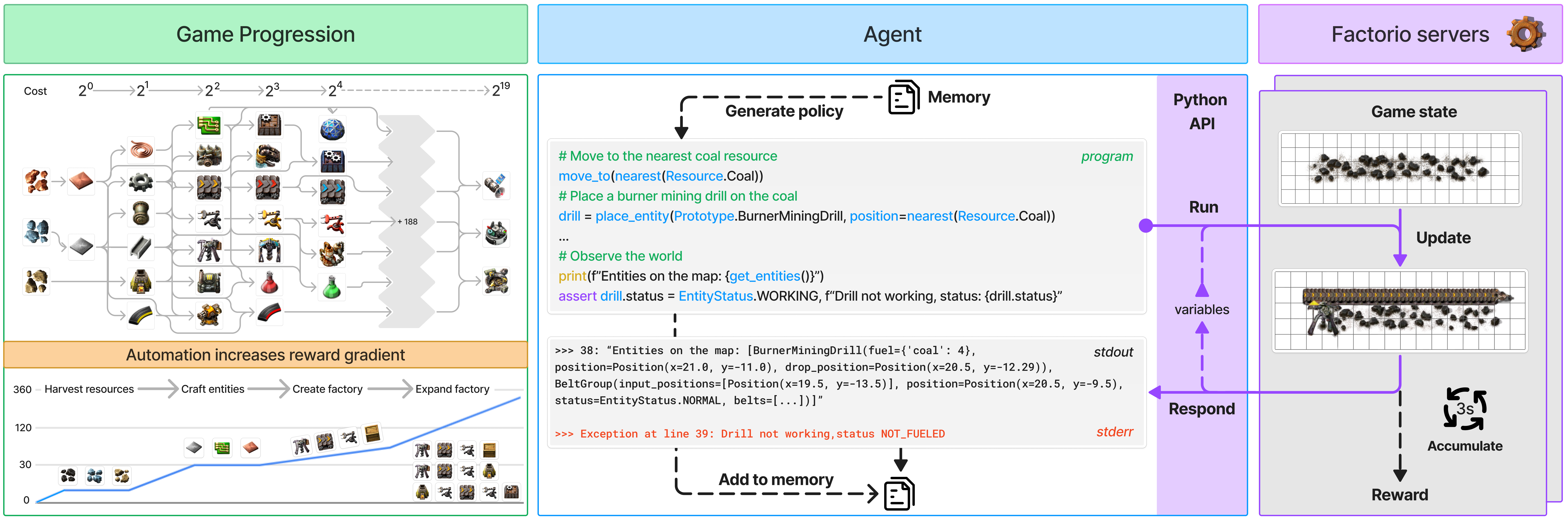}
\caption{\textbf{Illustration of the Factorio Learning Environment (FLE)}. FLE is based on the popular construction and management simulation game \textit{Factorio}. Left: The open-ended goal of the game is to create the largest factory possible. The game enables agents to invest in (an infinite number of) technological advances to produce more resources per second. Middle: Agents interact with the game by using an interactive Python Interpreter, where they take actions and print their observations in a Read-Eval-Print loop. By using the Python namespace, agents may store variables and define functions for later use. We provide a Python API to Factorio which allows direct interaction with the environment. Right: The agent may issue commands to the game server in order to interact with the environment (with associated time penalities), and receive a response as feedback. If the agents chooses, it may view its own production statistics.}
\label{fig:fle_setup}
\end{figure*}

Agents interact with the FLE via synthesizing Python programs to alter and observe the game state, using the tools included in the environment in a Read\hyph Eval\hyph Print Loop (REPL). This feedback loop mirrors the day-to-day workflow of human programmers, who write provisional code to probe how systems behave, interpret the results, then refine their mental model of the system. In this sense, the agent’s program acts as the cumulative representation of its current knowledge and strategies for managing the complex resource pipelines in Factorio.

We evaluate six frontier LLM models in this environment in an agentic setting. In our qualitative analysis, we study the agents capabilities for spatial reasoning, long-term planning, and error correction. Our results show that even the most advanced models struggle to coordinate more than six machines when automatically producing items with over three ingredients, even after 128 environmental interactions.

We summarise our contribution as follows:
\begin{itemize}
   \item The introduction of the Factorio Learning Environment, an agentic evaluation of long-term planning and resource management and allocation. 
   \item Evaluations of frontier models in FLE lab-play, a set of 24 controlled tasks requiring agents to build factories with increasing levels of complexity and scale. Claude-3.5-Sonnet (the strongest performing model) only completes 7/24 tasks and shows limitations in spatial planning in more complex objectives; demonstrating large head-room for performance.
   \item Evaluation of frontier models in the FLE open-play, an unbounded setting in a full Factorio game map. We find more capable agents (such as Claude-3.5-Sonnet and GPT-4o) invest heavily into technological research achieving quantitatively different slopes on a log-reward, log-step graph.   
   \item A qualitative analysis of the results across capabilities such as error-correction and long-term planning. We identify a gap in models' ability to perform intelligent error correction, iteratively build upon prior work and conduct exploration.
\end{itemize}
\section{Factorio Learning Environment}

Our main contribution is the release of an open-source framework, which includes i) a high-level Python API to Factorio, ii) a persistent coding environment for LLM agents to interact with the game through iterative program synthesis, and iii) a Python object model of game entities. 

The environment is procedurally generated, deterministic at runtime (set by a random seed) and is $4 \times 10^{12}$ square tiles in size. We provide a laboratory environment with accessible resources for benchmarking agents in a controlled setting.

\subsection{Environment Dynamics}
Factorio is a resource management and automation game in which players spawn on a world containing raw resources such as water, iron ore, and coal, and must orchestrate increasingly complex production and logistic chains to ultimately produce a rocket and (optionally) escape. The game contains over 200 entity types, with a technology tree that unlocks more efficient buildings, resource production chains and multiplicative throughput bonuses. 
Research enforces a steep resource progression, with late-game technologies such as the \texttt{rocket-silo} demanding 300 times more resources than early \texttt{automation} research\footnote{This progression approximately follows an unbounded geometric relationship between resource cost $C$ and research tier $N$ -- $C[N] = 1000 \times 2^{(N-1)}$}.

Player strategy and factory architecture evolves dramatically as technology progresses. The early game centres on manual crafting and basic automation, with factories daisy-chained together using direct insertion between machines. These designs are primarily constrained by manual building speed and resource gathering, favouring cheap, immediate solutions \hyph e.g \texttt{stone-furnaces} provide better returns on investment than \texttt{steel-furnaces} until fast-belt technology (\texttt{logistics-2}) is available. As they progress, players typically adopt main bus designs with centralized resource production and distribution, enabling more organized scaling of production. Late-game strategy shifts again, toward massive parallel construction and logistics networks, with factories ultimately evolving into distributed complexes connected by high-capacity train networks. These advanced stages emphasize space-efficient, high-throughput designs enabled by technologies like beacons and stack inserters, and are thousands of times larger than early game factories.

\begin{figure}
\label{fig:automated-mining}
\lstdefinestyle{pythoncode}{
    language=Python,
    basicstyle=\ttfamily,
    keywordstyle=\color{blue},
    stringstyle=\color{green!50!black},
    commentstyle=\color{gray},
    numbers=left,
    numberstyle=\tiny,
    numbersep=5pt,
    showstringspaces=false,
    breaklines=true,
    frame=single
}
\begin{lstlisting}[language=python]
# 1. Get iron patch and place mining drill
drill = place_entity(
    entity=Prototype.MiningDrill,
    position=nearest(Resource.IronOre)),
    direction=Direction.NORTH
)
# 2. Add output storage
chest = place_entity_next_to(
    entity=Prototype.IronChest,
    reference_position=drill.drop_position,
    direction=Direction.SOUTH
)
# 3. Verify automation chain and observe entities
assert drill.status == EntityStatus.WORKING
print(get_entities())
\end{lstlisting}
\caption{\textbf{Example of an FLE program} used to create a simple automated iron-ore miner. In step 1 the agent uses a query to find the nearest resources and place a mine. In step 3 the agent uses an assert statement to verify that its action was successful.}
\end{figure}

\subsection{Environment Interface}

Agents interact with FLE through a \textbf{REPL} (Read-Eval-Print-Loop) pattern, observing the current game state via previous program output streams, then generating and executing Python code to implement their intended actions, and finally returning useful feedback for the next iteration. 

Agents are provided with the Python standard library, and an API comprising methods designed to balance expressiveness with tractability (see Appendix~\ref{appendix:api}). These comprise 10 observation methods and 13 action methods. Observation methods (e.g \texttt{nearest}, \texttt{get\_entities}) retrieve information about the environment, and action methods (e.g \texttt{move\_to}, \texttt{craft\_entity}) modify the environment. 

\begin{figure*}[ht]
\includegraphics[width=1\textwidth]{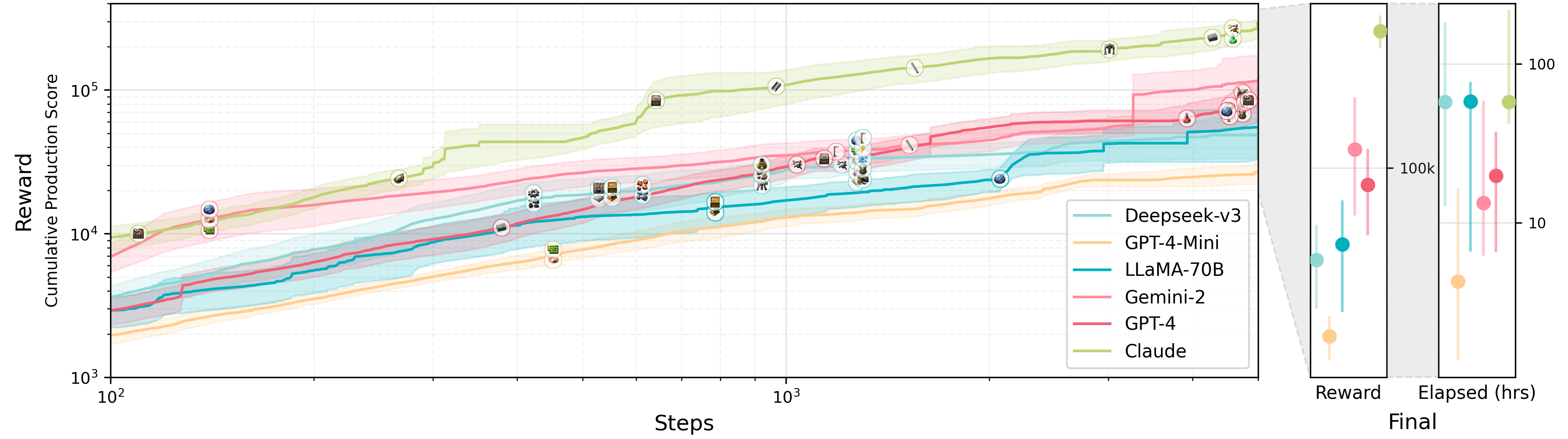}
\caption{\textbf{Models are differentiated by score in Open-Play.} Agents are given the instruction to \textit{build the biggest possible factory}. Left: We find that by evaluating PS against steps (server calls) we can clearly differentiate stronger models from weaker ones in a log/log projection. We overlay milestones, showing the first time the median agent was able to create a new type of entity. Right: We plot the final reward and elapsed game time after 5k steps. We find that while weaker models show promise early-game, they struggle to progress when automation and logistics are required. 
We report median and standard error over the independent runs.}
\label{fig:progression-steps}
\end{figure*}

Each method returns a typed object (e.g an \texttt{Inventory}) which can be stored as a variable in the Python namespace and referenced later in the episode. The namespace acts as an episodic symbolic memory system, and saved objects represent part of the environment at the moment of query, becoming stale as the game state evolves, requiring the agent to re-query when appropriate. This design enables agents to maintain complex state representations and build hierarchical abstractions as the factories scale. 

Agents observe \textbf{stdout} and \textbf{stderr} - the output streams of their program. Thus, agents may intentionally print relevant objects and computations to the output stream to construct observations. Selecting relevant attributes of objects enables token-efficient observation, e.g:
\begin{lstlisting}[language=Python]
# Observe the status of all Furnaces in-game
print([e.status for e in get_entities({Prototype.StoneFurnace})}
>>> 2: 'WORKING', 'WORKING', 'NEED_INGREDIENTS',
       'OUT_OF_FUEL'
\end{lstlisting}
Mistakes in the code or invalid operations raise typed exceptions with detailed context that is written to \textbf{stderr}. 
This enables agents to \textit{reactively} debug their programs after execution, and \textit{proactively} use runtime assertions during execution to self-verify their actions. 
Programs that take too long to execute are terminated, to prevent runaway control flows (e.g while True).

An environment ``step'' is a single submission to the Factorio server, which returns the stdout, stderr, rewards and in-game wall-clock time (see Figure~\ref{fig:fle_setup}). 

Agents are able to enhance their internal representation of the game state in 2 ways: (i) they can define utility functions for reuse throughout an episode, to encapsulate successful logic; and (ii) they can define classes in the namespace to better organize the data retrieved from the game.

\subsection{Reward Structure}
We use Factorio's built-in production tracking system, which enables us to define two complementary reward signals:

\textbf{Production Score (PS):} A continuous measure of economic activity based on the value of all items produced. This metric increases as agents refine raw ores into manufactured goods and create automatic factories. As production chain throughput scales exponentially, PS can vary by multiple orders of magnitude (a rocket launch requires $\approx
10^7$ raw resources). PS provides a naturally unbounded measure of performance, which is sensitive to increasing automation complexity. The game's price calculation system assigns higher value to items with more complex production chains, creating a reward structure that encourages sophisticated factory designs. For the full pricing system, see Appendix~\ref{appendix:economics}.

\textbf{Milestones}: A discrete set of achievements for producing novel item types (e.g.\ building an \texttt{inserter} for the first time, assembling \texttt{electronic-circuits}, etc.) and researching technologies. This captures both the diversity of an agent’s exploration across Factorio’s tech tree, and what level of item complexity they were able to achieve. As Factorio supports researching an infinite technologies (with multiplicative bonuses), milestones can be used to measure performance at all levels of capability.

\subsection{Implementation Details}
The FLE comprises a Python client and Lua server communicating synchronously via RCON over TCP\footnote{Roughly 80k LoC in total}. The client provides the stateful environment interface and APIs, while the server manages game state execution in the official Factorio multiplayer server. The server can be run in headless mode for efficient parallelization. The object model represents most early to late-game entities (detailed in Appendix \ref{appendix:api}). FLE is compatible with v1.110 of Factorio, and requires a single purchased game license, as each server must be ``activated'' by any official client at startup. FLE is also easily extensible by the community. Designing new tools requires implementing a client-side controller (Python) and a server-side action (Lua) which will automatically load and update the API schema for subsequent agent runs.

We benchmark the Factorio Learning Environment on a MacBook Pro M4 with 128GB RAM. The headless server achieved the highest throughput, processing an average of 218 operations per second across core API functions, with peak performance of 603 ops/sec for basic operations like crafting. The Python interpreter introduces approximately 3x overhead, reducing average throughput to 68 ops/sec. Complex spatial operations (\texttt{connect\_entities}) are consistently the slowest at 25-48 ops/sec due to pathfinding requirements. Basic inventory operations (\texttt{craft\_item}, \texttt{extract\_item}) achieve highest throughput at 276-545 ops/sec. The headless configuration provides a 1.75x speed-up over the game client (see Figure~\ref{fig:benchmark-comparison}). We make the environment publicly available \footnote{https://github.com/JackHopkins/factorio-learning-environment} .

\section{Experiments}
To evaluate agent capabilities in FLE, we introduce two settings and a novel agent scaffolding.

\subsection{Settings}
\textbf{Lab Play} - requires the agent to create a factory with a specific production throughput in the constrained lab environment. These tasks are designed to evaluate the capabilities of an agent to create automatic structures in an open-ended manner, requiring creativity, spatial understanding of the map and long-horizon planning.

\begin{figure*}[t]
\includegraphics[width=1\textwidth]{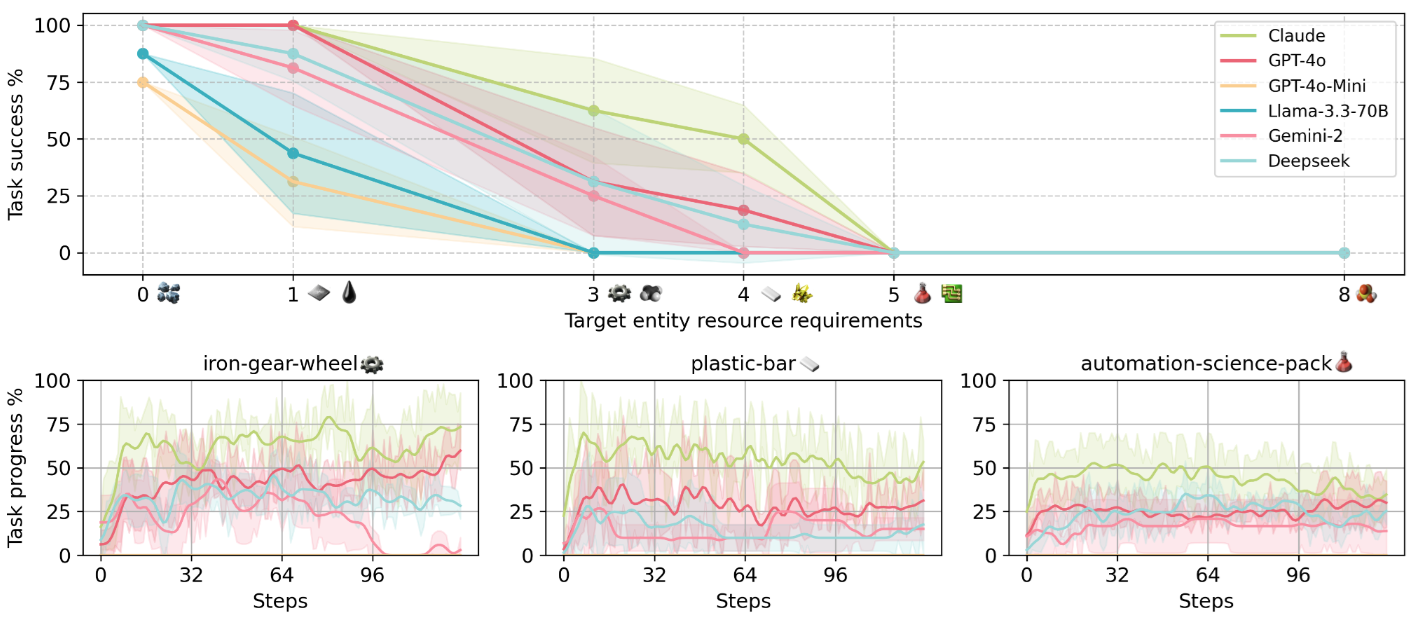}
\caption{\textbf{Agents are unable to consistently build complex and efficient factories in Lab-Play}. Top: We measure the mean and standard deviation of task success rates across the first 8 complexity levels and task progress (percentage of target ingredients and its sub-ingredients agents factories produce at each time-step) in three tasks of increasing difficulty. We observe a clear decrease in average task success rates as the crafting complexity of the target entity increases. Bottom: In harder tasks, agents show trends of initial rapid progress followed by stagnation or decrease. This is due to agents being unable to scale up initial production or add new sections to factories required to successfully reach the target production levels and often breaking existing structures during the process. The lack of consistent progress is also observed through the large variance in mean task progress across runs.}
\label{fig:results-lab-trendlines}
\end{figure*}

We task agents to build production lines of 24 distinct target entities of increasing complexity, starting from a single resource mine requiring at most 2 machines (making \texttt{iron-ore}) to a late game entity requiring the coordination of close to 100 machines (making \texttt{utility-science-pack}). The target entities cover items from early to late game and the agent must use a wide variety of machines present in Factorio (drills, furnaces, assembling machines, oil refineries, chemical plants). As the task difficulty naturally increases with resource requirements this provides a measure of the complexity agents are capable of creating in a limited number of steps. All tasks start with an inventory containing sufficient entities to complete the task and all research is unlocked for the agent in \textit{lab-play}. Additional information are in Appendix \ref{lab-play}.

Each task runs a trajectory of \textit{128} API calls. After every agent step, the throughput of the created structure is evaluated throughout a 60 second holdout period in-game, and the task is deemed completed if the throughput of the structure is above the target throughput at any step $i$. The target throughput is \textit{16} for solid items (for instance electronic circuit, military science pack, plastic bar) and \textit{250} for fluids (for instance petroleum gas, lubricant, heavy oil) during the holdout period. We report the mean and std of each task success with 4 runs per task.

\textbf{Open Play} - In addition to the structured \emph{lab-play} tasks, we evaluate each agent in a purely open-ended setting. The agents spawn into a procedurally generated world with unbounded space and resources, and are tasked to ``build the largest factory possible'', allowing the agents to decide how best to advance in the game. To progress long-term, agents must show proficient long-term goal-setting, entity and resource planning and spatial reasoning capabilities when creating automation structures. Agents must be capable of using the API, querying the environment for unknown information and reasoning over observations to plan successfully. 

We use two metrics to evaluate progress in the game: \textit{Production Score} (PS) and \textit{Milestones}. While the PS acts as the reward and is affected by exploitation, milestones give an overview of how much of the game and technology tree the agent has explored.

Each agent plays until the maximum trajectory length of 5000 is reached. After every agent step, the production throughput is tracked and reward computed. We execute 8 independent runs for each agent, and report the median.

\subsection{Agent Scaffolding}
We consider a simple step-by-step prompting approach as a baseline implementation for agents to interact with the environment. The input prompt of the agent consists of the API schema \textit{A}, a guide \textit{G} describing common patterns with code examples and the memory \textit{M} of the agent consisting of past policies with environment observations. A detailed description for the guide, API schema and an example memory state is brought out in Appendix \ref{agent-scaffolding}.
Given the inputs, the agent is tasked to identify the most useful next step and generate the Python policy \textit{$P$} that carries out actions in the environment to achieve the step. The policy is executed in the environment and added to the memory $M$ with the environment observations (\textbf{stdout}) and error messages (\textbf{stderr}). The updated memory $M$ is then used as input to generate the next policy and enables the agent to gather information from the environment and use observations to guide future step generation.

\textbf{Memory} - At every policy generation step the agent uses information from the memory $M$. Memory consists of past policies and their respective environment observations (stdout and stderr after every policy execution). To limit the memory token count, past observations and policies that are further than 32 steps in the past are summarised into a report of 1024 tokens. This allows the agent to execute arbitrarily long traces in the environment without unreasonably large memory input token requirements.

\textbf{Language Models} - We evaluate state-of-the-art closed source models including Claude 3.5-Sonnet \cite{anthropic2024claude35}, GPT-4o and GPT-4o-Mini \cite{openai2024gpt4technicalreport}, Deepseek-v3 \cite{deepseekai2025deepseekv3technicalreport} and Gemini-2-Flash \cite{geminiteam2024geminifamilyhighlycapable}. We also evaluate Llama-3.3-70B-Instruct \cite{metaai2024llama}. Each model is sampled at temperature $0.5$. Model timestamps are in Appendix~\ref{section:model-access}.


\section{Results}

We analyse agent performance during \textit{open-play} and \textit{lab-play}, and observe common patterns amongst trajectories from both settings. We report experimental costs in Table~\ref{tab:token-costs}.

\begin{figure*}[ht]
\includegraphics[width=1\textwidth]{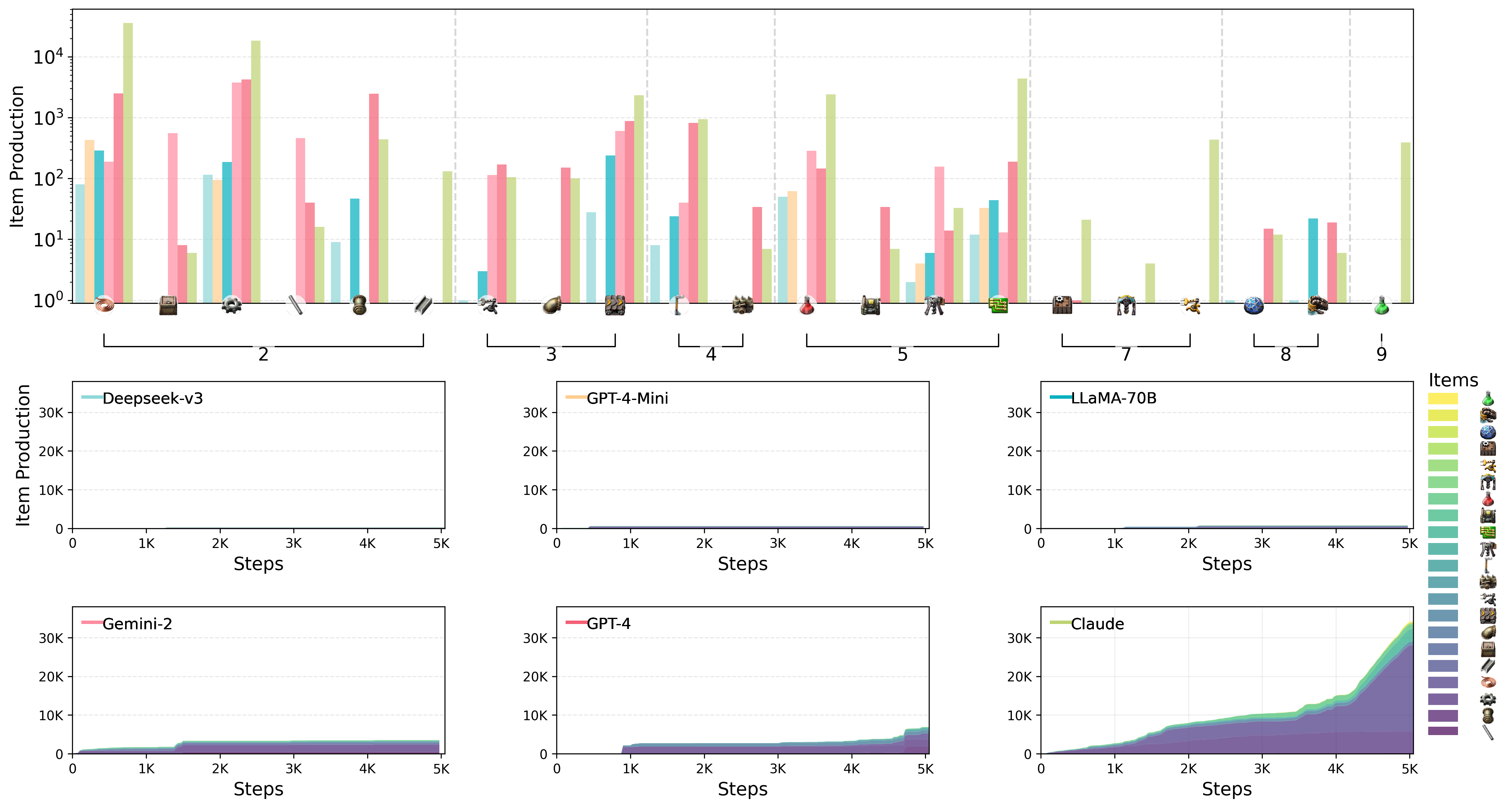}
\caption{\textbf{Open-ended challenges highlight differences in objective setting and general capability.} We illustrate the rates at which various models produce items with multiple antecedent ingredients in the \textit{open-play} setting. Claude 3.5-Sonnet immediately begins complex crafting and invests in research, ultimately unlocking \texttt{electric-mining-drills} at step 3k, the deployment of which boosts production of \texttt{iron-plate} thereafter. Less advanced models, like GPT-4o-Mini, produce insignificant quantities of multi-ingredient items. Deepseek produced fewer complex resources in open-play than its lab-play performance would suggest, indicating weaker objective-setting compared to general capability in-game.  }
\label{fig:results-open-play-complex-achievements}
\end{figure*}
\textbf{Insight 1: Coding agents perform better in the FLE.} We find that stronger coding agents achieve higher production scores across both settings. In \textit{open-play}, Claude outperforms other models in both median PS (293 206) and milestone count (28), surpassing the early-game resource extraction phase and partially investing in technology research - constructing and powering a \texttt{lab}, dedicating production to \texttt{science-packs} and unlocking electric \texttt{inserters}, (see Figure~\ref{fig:results-open-play-complex-achievements}). In comparison Llama-3.3-70B (54 998 PS, 26 milestones) made initial progress but did not develop production lines of \textgreater3 entities and struggled with both creating complex structures and scaling up existing production. Similarly in \textit{lab-play}, Claude performed the best, managing to create automatic structures typically seen in Factorio's early game; specifically, compact drilling lines coordinating 10+ machines across up to four factory sections (see table \ref{tab:lab-play-performance}). In comparison,  Llama-3.3-70B is capable of only creating the most trivial of factories. 

\textbf{Insight 2: Agents lack spatial reasoning and are unable to iteratively improve on factories.} A key characteristic for success in \textit{open-play} and \textit{lab-play} involves iteratively combining multiple factory sections to create complex production lines. In \textit{open-play}, while Claude was able to scale up automation from the early-game, GPT-4o, GPT-4o-Mini and Llama-3.3-70B typically succeeded only at maintaining a small number of production lines. In lab-play, it can be seen how the success rate of tasks decreases proportionally to the increase in crafting recipe complexity of the target entity (see Figure \ref{fig:results-lab-trendlines}). For instance, creation of automation science packs requires multiple mining and smelting sections (\texttt{iron-plate} and \texttt{copper-plate}), a \texttt{iron-gear-wheel} assembly section, \texttt{automation-science-pack} section and a \texttt{steam-engine}. While agents are able to make initial progress in this task by creating electricity setups and plate production lines, they are unable to improve on the factory and add the required assembly sections. Frequent failure cases were trying to place entities too close or on-top of each other, not leaving room for connections or incorrect placement of inserters. These are all limitations in spatial reasoning and result in agents only being able to consistently create production lines for low complexity items and low overall performance in \textit{lab-play} tasks (see table \ref{tab:lab-play-performance}).

\begin{table}
\centering
\begin{tabular}{l|c}
\toprule
Model & Lab-play task success rate (\%)\\
\midrule
Claude & 21.9 $\pm$ 1.3\\
 GPT-4o & 16.6 $\pm$ 1.4\\
 Deepseek-v3 & 15.1 $\pm$ 1.7\\
 Gemini-2 & 13.0 $\pm$ 1.3\\
 Llama-3.3-70B & 6.3 $\pm$ 1.0\\
 GPT-4o-Mini & 5.2 $\pm$ 0.6\\
\bottomrule
\end{tabular}
\caption{\textbf{Stronger coding models achieve higher task success rates in Lab-Play}. We observe a correlation between coding and reasoning abilities of base models and the task success rates, where the stronger models have higher success rates in lab-play tasks. Claude, GPT-4o, Deepseek and Gemini-2 are able to only pass simpler tasks requiring the construction of factories consisting up to 3 sections while Llama-3.3-70B and GPT-4o-Mini succeeded in tasks requiring single-section factories. All models struggled with the increasing scale and complexity requirements for tasks resulting in low overall success rates.}
\label{tab:lab-play-performance}
\end{table}
\textbf{Insight 3: Agents use the API in different ways.} We evaluate trajectories with automatic checkers to evaluate how successful models are at using the FLE API. We find that models exhibit different coding styles, with GPT-4o using more assert checks in within their code than Claude 3.5. Conversely GPT-4o uses significantly fewer prints. These suggest models use very different approaches to explore and engage with the FLE. Using prints suggests being uncertain of state, and exploring new areas, whereas assert statements are likely used to clarify existing knowledge (see Table \ref{tab:code}).

\textbf{Insight 4: Planning is essential to open-play performance}. 
In \textit{open-play}, agents are given an open-ended goal and need to create meaningful sub-objectives themselves to make long-term progress. We observe that agents often set short-sighted objectives, for instance manually crafting a large quantity of singular entities without a long-term plan (Gemini-2.0 manually crafted 300+ wooden chests over 100 steps), not significantly investing into research (except for Claude) or creating small individual factories as opposed to scaling up existing production. This creates a discrepancy between \textit{lab-play} and \textit{open-play} results where in lab-play Gemini-2 and Deepseek show capabilities in creating early-game automation (see Figure \ref{fig:results-lab-trendlines}) but rarely attempt creating factories in open-play, resulting in poor complex entity crafting statistics (See Figure \ref{fig:results-open-play-complex-achievements}) and lower production scores.  

\begin{table}[t]
\centering
\scriptsize
\begin{tabular}{lrrrr|rrr}
\toprule
Model & L & P\% & A\% & F\% & AF\% & C\% & En\% \\
\midrule
Claude-3.5 & 65 & 43.3 & 2.0 & 50.6 & 0 & 3 & 97 \\
GPT-4o & 81 & 10.3 & 12.8 & 10.2 & 2 & 12 & 86 \\
DeepSeek-v3 & 37 & 25.4 & 13.9 & 25.3 & 0 & 2 & 98 \\
Gemini-2 & 133 & 16.2 & 0.0 & 16.6 & 1 & 46 & 53 \\
Llama-3.3-70B & 38 & 23.9 & 12.9 & 23.7 & 0 & 24 & 76 \\
GPT-4o-Mini & 77 & 36.0 & 0.0 & 31.6 & 15 & 6 & 79 \\
\bottomrule
\end{tabular}
\label{tab:code}
\caption{\textbf{Models exhibit contrasting coding styles}: Analysis of code submitted by different language models, showing average lines per program (L), percentage of lines that were print statements (P\%), percentage of lines that were assertions (A\%), and percentage of programs that failed (F\%). For programs that failed by error type we track the proportion of assertion fails (AF\%), code errors (C\%), and environment errors (En\%). Claude-3.5 favours a REPL approach with high print usage and failure rates, while GPT-4 opts for defensive programming with assertive validation and fewer resulting environmental errors. Gemini-2 produces the longest programs (133 lines on average) but makes the most code errors (e.g accessing non-existent variables).}
\end{table}

\textbf{Insight 5: Agents which invest in technological progression achieve much higher PS}. Investing into technology progress in open-play is a trade-off, where agents incur a short-term resource cost to unlock items enabling long-term higher throughput. Although research is crucial for creating higher efficiency factories, only Claude consistently invests resources into researching new technologies in open-play. The result can be seen from step 3k, where Claude starts deploying \texttt{electric-mining-drills} and PS grows by a factor of 1.5x (from 200k to 300k), in Figure~\ref{fig:results-open-play-complex-achievements}.

\textbf{Insight 6: Agents fall into degenerate debug loops.} A critical component for successful runs was an agents' ability to interact to previous error logs and carry out error correction. In \textit{lab-play}, in successful task completions, $56\%$ of steps resulted in program execution errors (from which agents recovered), and in \textit{open-play} ranged from $29.7\%$ to $76.4\%$. Claude, GPT-4o and Deepseek were capable of simpler error correction when incorrectly using the API or when crafting entities. Anecdotally, the agents were not proficient at debugging complex environments. For instance, when debugging non-working structures or factories where the throughput was not at expected levels, agents often focused on whether all singular entities were working but did not investigate whether the topology of the whole structure was correct. In \textit{lab-play}, this limitation is illustrated by the frequent decrease of task performance across steps in Figure \ref{fig:results-lab-trendlines} where the agents broke existing working structures due to incorrectly identifying the root-cause of problems. Agents often fell into a loop of greedily repeating the same fix rather than exploring additional potential sources of the problem. This can be seen in the flatline behaviour during open-play in Figure \ref{fig:progression-steps} with no PS progression. For instance, in one run GPT-4o used the same API method incorrectly for 78 contiguous steps (from Step 120), receiving identical error message each time. On two occasions, GPT-4o-Mini simply gave up and repeatedly asked to be reset - see Appendix~\ref{section:4o-mini-failure}.


These limitations show the difficulty of FLE and that state-of-the-art LLMs---even with REPL-style feedback and extensive prompting---are still in the early stages of mastering large-scale, open-ended factory design. 

\section{Related Work}

Games have long served as fundamental benchmarks for artificial intelligence research, providing standardized environments with clear metrics, rich observational data, and natural difficulty gradients.~\citep{DeepBlue, AlphaGo, berner2019dota}.

Recent work has explored using LLMs as game-playing agents. Environments like ALFWorld~\cite{alfworld} combine language understanding with embodied tasks, while MineDojo~\cite{MineDojo} leverages Minecraft as a sandbox for testing general-purpose agents through 3,000+ diverse tasks spanning survival, harvesting, and creative building. While these environments excel at evaluating breadth of capabilities, their fundamentally linear progression systems limit their ability to differentiate highly capable agents. Even with rich task suites, their resource requirements remain relatively modest compared to Factorio's exponential scaling.

Many benchmarks exist for agentic coding such as \citep{jimenez2023swe,hendrycks2measuring}, which evaluate Python against a stateful system. These benchmarks often involve fitting specific test conditions, or task descriptions. In comparison, our win condition is open-ended, requiring multi-step planning and resulting in thousands of submissions; requiring much longer contexts than other benchmarks.

Management simulation games like OpenTTD~\cite{OpenTTD} have explored aspects of resource optimization, but lack precise mechanics and exponential scaling that would enable quantitative measurement of small improvements in agent capabilities. Text-based environments like Jericho~\cite{jericho} test language understanding through interactive fiction, but lack the spatial reasoning and timing requirements inherent in factory design. 

Factorio has seen prior research interest for closed-domain settings ~\cite{reid2021factorygrowautomationfactorio}, with a focus on integer programming models, meta-heuristics and evolutionary reinforcement learning to tackle logistics challenges. We build on this foundation to offer a standardized text-based interface for learning agents to solve open-ended challenges in long-term planning, spatial reasoning and factory optimization.

\section{Limitations, Future Work \& Conclusion}

In this work, we introduce the Factorio Learning Environment (FLE), a novel open-ended framework for evaluating the capabilities of agents in an open-ended environment. 

A major concern for any environment benchmark is reward hacking \citep{clark2016faulty, skalse2022defining}. In our setting this could involve two main attack surfaces: either through Python API (as seen within \citet{denison2024sycophancy}) or within the Factorio game-engine itself. During our evaluations, while we observed no direct examples of reward hacking of either interface, we did observe that the agent was able to occasionally trigger resetting the Factorio gamestate; That said, we applied little optimisation pressure on agents - we highlight this as a caution for those developing agents on FLE.  

Secondly, unlike the base game of Factorio, which uses mouse and screen, our approach involves a Python interface. Whilst the authors were, in equivalent game time, able to outperform frontier agents, it is unclear if achieving end-game goals (e.g. escape the world or build rockets) is achievable to humans using only an API in a reasonable time-frame. We did however prove that each step in the chain to launch a rocket was achievable from the previous step and that all tasks in \textit{lab-play} can be completed.

Even without human baselines, we believe that FLE is a useful benchmark, as the comparative scores between agents still informs us of their relative ability at planning, spatial reasoning and resource management. 

Currently, some mid and late-game entities are not explicitly modelled in FLE (for instance trains, logistics robots and agent-programmable circuit networks). This is appropriate given the state of evaluated models. As FLE is open-source, we are committed to release a complete object model with first-class support for all entities.

Although all our current experiments use single-agent interaction, Factorio inherently supports multi-player games in both cooperative and competitive scenarios. For instance, multiple agents could share a base, coordinating research and logistics, or they could spawn in distant regions and compete for finite resources (e.g., high-yield iron patches). 

Through our evaluation, we demonstrate that even state-of-the-art agents struggle with the coordination and optimization challenges inherent in simple automation and logistical tasks. The limitations we observed in spatial reasoning, long-term planning, and intelligent error correction highlight gaps in capabilities of foundation language models in novel environments.
The unbounded nature of FLE provides a benchmark that will remain relevant even as progress in LLMs continues to advance. Unlike traditional benchmarks that are rapidly saturated by progress in AI research, FLE's exponentially scaling reward system and requirement for capabilities across multiple areas create a natural curricula that can meaningfully differentiate between increasingly performant models.




\section*{Impact Statement}

The Factorio Learning Environment provides a novel, open-source testbed for exploring advanced AI behavior in a complex, open-ended setting. It enables rigorous empirical study of the instrumental convergence hypothesis and other AI safety concerns. Researchers can measure how agents balance resource acquisition, territorial expansion, and defense, offering early indicators of convergent behaviors and potential risks. Moreover, Factorio’s diverse automation challenges allow systematic comparisons of agent architectures and interventions, fostering reproducible research at scale. We hope this platform accelerates empirical safety investigations and strengthens the connection between theoretical predictions and real-world AI behavior.

\nocite{langley00}

\bibliography{example_paper}

\begin{thebibliography}{31}
\providecommand{\natexlab}[1]{#1}
\providecommand{\url}[1]{\texttt{#1}}
\expandafter\ifx\csname urlstyle\endcsname\relax
  \providecommand{\doi}[1]{doi: #1}\else
  \providecommand{\doi}{doi: \begingroup \urlstyle{rm}\Url}\fi

\bibitem[Ope()]{OpenTTD}
{OpenTTD: Transport simulation game}.
\newblock \url{https://www.openttd.org/}.
\newblock Accessed: 2025-01-07.

\bibitem[{Anthropic}(2024)]{anthropic2024claude35}
{Anthropic}.
\newblock Claude 3.5 sonnet: Enhanced intelligence and versatility, 2024.
\newblock URL \url{https://www.anthropic.com/news/claude-3-5-sonnet}.
\newblock 20241022-v2:0 version.

\bibitem[Berner et~al.(2019)Berner, Brockman, Chan, Cheung, Debiak, Dennison, Farhi, Fischer, Hashme, Hesse, et~al.]{berner2019dota}
Berner, C., Brockman, G., Chan, B., Cheung, V., Debiak, P., Dennison, C., Farhi, D., Fischer, Q., Hashme, S., Hesse, C., et~al.
\newblock Dota 2 with large scale deep reinforcement learning.
\newblock \emph{arXiv preprint arXiv:1912.06680}, 2019.

\bibitem[Campbell et~al.(2002)Campbell, Hoane, and Hsu]{DeepBlue}
Campbell, M., Hoane, A.~J., and Hsu, F.-h.
\newblock {Deep Blue}.
\newblock \emph{Artificial Intelligence}, 134\penalty0 (1--2):\penalty0 57--83, 2002.
\newblock \doi{10.1016/S0004-3702(01)00129-1}.

\bibitem[Chen et~al.(2021)Chen, Tworek, Jun, Yuan, Pinto, Kaplan, Edwards, Burda, Joseph, Brockman, et~al.]{chen2021evaluating}
Chen, M., Tworek, J., Jun, H., Yuan, Q., Pinto, H. P. D.~O., Kaplan, J., Edwards, H., Burda, Y., Joseph, N., Brockman, G., et~al.
\newblock Evaluating large language models trained on code.
\newblock \emph{arXiv preprint arXiv:2107.03374}, 2021.

\bibitem[Clark \& Amodei(2016)Clark and Amodei]{clark2016faulty}
Clark, J. and Amodei, D.
\newblock Faulty reward functions in the wild.
\newblock \emph{Internet: https://blog. openai. com/faulty-reward-functions}, 2016.

\bibitem[Cobbe et~al.(2021)Cobbe, Kosaraju, Bavarian, Chen, Jun, Kaiser, Plappert, Tworek, Hilton, Nakano, et~al.]{cobbe2021training}
Cobbe, K., Kosaraju, V., Bavarian, M., Chen, M., Jun, H., Kaiser, L., Plappert, M., Tworek, J., Hilton, J., Nakano, R., et~al.
\newblock Training verifiers to solve math word problems.
\newblock \emph{arXiv preprint arXiv:2110.14168}, 2021.

\bibitem[DeepSeek-AI et~al.(2025)DeepSeek-AI, Liu, Feng, Xue, Wang, Wu, Lu, Zhao, Deng, Zhang, Ruan, Dai, Guo, Yang, Chen, Ji, Li, Lin, Dai, Luo, Hao, Chen, Li, Zhang, Bao, Xu, Wang, Zhang, Ding, Xin, Gao, Li, Qu, Cai, Liang, Guo, Ni, Li, Wang, Chen, Chen, Yuan, Qiu, Li, Song, Dong, Hu, Gao, Guan, Huang, Yu, Wang, Zhang, Xu, Xia, Zhao, Wang, Zhang, Li, Wang, Zhang, Zhang, Tang, Li, Tian, Huang, Wang, Zhang, Wang, Zhu, Chen, Du, Chen, Jin, Ge, Zhang, Pan, Wang, Xu, Zhang, Chen, Li, Lu, Zhou, Chen, Wu, Ye, Ye, Ma, Wang, Zhou, Yu, Zhou, Pan, Wang, Yun, Pei, Sun, Xiao, Zeng, Zhao, An, Liu, Liang, Gao, Yu, Zhang, Li, Jin, Wang, Bi, Liu, Wang, Shen, Chen, Zhang, Chen, Nie, Sun, Wang, Cheng, Liu, Xie, Liu, Yu, Song, Shan, Zhou, Yang, Li, Su, Lin, Li, Wang, Wei, Zhu, Zhang, Xu, Xu, Huang, Li, Zhao, Sun, Li, Wang, Yu, Zheng, Zhang, Shi, Xiong, He, Tang, Piao, Wang, Tan, Ma, Liu, Guo, Wu, Ou, Zhu, Wang, Gong, Zou, He, Zha, Xiong, Ma, Yan, Luo, You, Liu, Zhou, Wu, Ren, Ren, Sha, Fu, Xu, Huang, Zhang, Xie, Zhang, Hao,
  Gou, Ma, Yan, Shao, Xu, Wu, Zhang, Li, Gu, Zhu, Liu, Li, Xie, Song, Gao, and Pan]{deepseekai2025deepseekv3technicalreport}
DeepSeek-AI, Liu, A., Feng, B., Xue, B., Wang, B., Wu, B., Lu, C., Zhao, C., Deng, C., Zhang, C., Ruan, C., Dai, D., Guo, D., Yang, D., Chen, D., Ji, D., Li, E., Lin, F., Dai, F., Luo, F., Hao, G., Chen, G., Li, G., Zhang, H., Bao, H., Xu, H., Wang, H., Zhang, H., Ding, H., Xin, H., Gao, H., Li, H., Qu, H., Cai, J.~L., Liang, J., Guo, J., Ni, J., Li, J., Wang, J., Chen, J., Chen, J., Yuan, J., Qiu, J., Li, J., Song, J., Dong, K., Hu, K., Gao, K., Guan, K., Huang, K., Yu, K., Wang, L., Zhang, L., Xu, L., Xia, L., Zhao, L., Wang, L., Zhang, L., Li, M., Wang, M., Zhang, M., Zhang, M., Tang, M., Li, M., Tian, N., Huang, P., Wang, P., Zhang, P., Wang, Q., Zhu, Q., Chen, Q., Du, Q., Chen, R.~J., Jin, R.~L., Ge, R., Zhang, R., Pan, R., Wang, R., Xu, R., Zhang, R., Chen, R., Li, S.~S., Lu, S., Zhou, S., Chen, S., Wu, S., Ye, S., Ye, S., Ma, S., Wang, S., Zhou, S., Yu, S., Zhou, S., Pan, S., Wang, T., Yun, T., Pei, T., Sun, T., Xiao, W.~L., Zeng, W., Zhao, W., An, W., Liu, W., Liang, W., Gao, W., Yu, W., Zhang, W.,
  Li, X.~Q., Jin, X., Wang, X., Bi, X., Liu, X., Wang, X., Shen, X., Chen, X., Zhang, X., Chen, X., Nie, X., Sun, X., Wang, X., Cheng, X., Liu, X., Xie, X., Liu, X., Yu, X., Song, X., Shan, X., Zhou, X., Yang, X., Li, X., Su, X., Lin, X., Li, Y.~K., Wang, Y.~Q., Wei, Y.~X., Zhu, Y.~X., Zhang, Y., Xu, Y., Xu, Y., Huang, Y., Li, Y., Zhao, Y., Sun, Y., Li, Y., Wang, Y., Yu, Y., Zheng, Y., Zhang, Y., Shi, Y., Xiong, Y., He, Y., Tang, Y., Piao, Y., Wang, Y., Tan, Y., Ma, Y., Liu, Y., Guo, Y., Wu, Y., Ou, Y., Zhu, Y., Wang, Y., Gong, Y., Zou, Y., He, Y., Zha, Y., Xiong, Y., Ma, Y., Yan, Y., Luo, Y., You, Y., Liu, Y., Zhou, Y., Wu, Z.~F., Ren, Z.~Z., Ren, Z., Sha, Z., Fu, Z., Xu, Z., Huang, Z., Zhang, Z., Xie, Z., Zhang, Z., Hao, Z., Gou, Z., Ma, Z., Yan, Z., Shao, Z., Xu, Z., Wu, Z., Zhang, Z., Li, Z., Gu, Z., Zhu, Z., Liu, Z., Li, Z., Xie, Z., Song, Z., Gao, Z., and Pan, Z.
\newblock Deepseek-v3 technical report, 2025.
\newblock URL \url{https://arxiv.org/abs/2412.19437}.

\bibitem[Denison et~al.(2024)Denison, MacDiarmid, Barez, Duvenaud, Kravec, Marks, Schiefer, Soklaski, Tamkin, Kaplan, et~al.]{denison2024sycophancy}
Denison, C., MacDiarmid, M., Barez, F., Duvenaud, D., Kravec, S., Marks, S., Schiefer, N., Soklaski, R., Tamkin, A., Kaplan, J., et~al.
\newblock Sycophancy to subterfuge: Investigating reward-tampering in large language models.
\newblock \emph{arXiv preprint arXiv:2406.10162}, 2024.

\bibitem[Fan et~al.(2022)Fan, Xie, Shi, Sadat, Tan, Gong, Liang, and Huang]{MineDojo}
Fan, L., Xie, A., Shi, W., Sadat, A., Tan, X., Gong, W., Liang, J., and Huang, D.-A.
\newblock {MineDojo: Building Open-Ended Embodied Agents with Internet-Scale Knowledge}.
\newblock In \emph{Advances in Neural Information Processing Systems (NeurIPS)}, 2022.
\newblock URL \url{https://arxiv.org/abs/2210.14168}.

\bibitem[Guss et~al.(2019)Guss, Houghton, Topin, Wang, Codel, Veloso, and Salakhutdinov]{guss2019minerl}
Guss, W.~H., Houghton, B., Topin, N., Wang, P., Codel, C., Veloso, M., and Salakhutdinov, R.
\newblock Minerl: A large-scale dataset of minecraft demonstrations.
\newblock \emph{arXiv preprint arXiv:1907.13440}, 2019.

\bibitem[Hausknecht et~al.(2019)Hausknecht, Fulda, Tien, Murdock, Yarnell, and Stone]{jericho}
Hausknecht, M., Fulda, N., Tien, E., Murdock, J., Yarnell, D., and Stone, P.
\newblock {Jericho: A new RL environment for Interactive Fiction}.
\newblock \url{https://github.com/microsoft/jericho}, 2019.
\newblock Accessed: 2025-01-07.

\bibitem[Hendrycks et~al.()Hendrycks, Basart, Kadavath, Mazeika, Arora, Guo, Burns, Puranik, He, Song, et~al.]{hendrycks2measuring}
Hendrycks, D., Basart, S., Kadavath, S., Mazeika, M., Arora, A., Guo, E., Burns, C., Puranik, S., He, H., Song, D., et~al.
\newblock Measuring coding challenge competence with apps.
\newblock In \emph{Thirty-fifth Conference on Neural Information Processing Systems Datasets and Benchmarks Track (Round 2)}.

\bibitem[Hendrycks et~al.(2021)Hendrycks, Burns, Kadavath, Arora, Basart, Tang, Song, and Steinhardt]{hendrycks2021measuring}
Hendrycks, D., Burns, C., Kadavath, S., Arora, A., Basart, S., Tang, E., Song, D., and Steinhardt, J.
\newblock Measuring mathematical problem solving with the math dataset.
\newblock \emph{arXiv preprint arXiv:2103.03874}, 2021.

\bibitem[Jimenez et~al.(2023)Jimenez, Yang, Wettig, Yao, Pei, Press, and Narasimhan]{jimenez2023swe}
Jimenez, C.~E., Yang, J., Wettig, A., Yao, S., Pei, K., Press, O., and Narasimhan, K.
\newblock Swe-bench: Can language models resolve real-world github issues?
\newblock \emph{arXiv preprint arXiv:2310.06770}, 2023.

\bibitem[Kambhampati et~al.(2024)Kambhampati, Valmeekam, Guan, Stechly, Verma, Bhambri, Saldyt, and Murthy]{kambhampati2024llms}
Kambhampati, S., Valmeekam, K., Guan, L., Stechly, K., Verma, M., Bhambri, S., Saldyt, L., and Murthy, A.
\newblock Llms can't plan, but can help planning in llm-modulo frameworks.
\newblock \emph{arXiv preprint arXiv:2402.01817}, 2024.

\bibitem[K{\"u}ttler et~al.(2020)K{\"u}ttler, Nardelli, Miller, Raileanu, Selvatici, Grefenstette, and Rockt{\"a}schel]{kuttler2020nethack}
K{\"u}ttler, H., Nardelli, N., Miller, A., Raileanu, R., Selvatici, M., Grefenstette, E., and Rockt{\"a}schel, T.
\newblock The nethack learning environment.
\newblock \emph{Advances in Neural Information Processing Systems}, 33:\penalty0 7671--7684, 2020.

\bibitem[Liu et~al.(2023)Liu, Yu, Zhang, Xu, Lei, Lai, Gu, Ding, Men, Yang, et~al.]{liu2023agentbench}
Liu, X., Yu, H., Zhang, H., Xu, Y., Lei, X., Lai, H., Gu, Y., Ding, H., Men, K., Yang, K., et~al.
\newblock Agentbench: Evaluating llms as agents.
\newblock \emph{arXiv preprint arXiv:2308.03688}, 2023.

\bibitem[{MetaAI}(2024)]{metaai2024llama}
{MetaAI}.
\newblock Llama 3.3, 2024.
\newblock URL \url{https://www.llama.com/docs/model-cards-and-prompt-formats/llama3_3/}.
\newblock Accessed: 2025-01-25.

\bibitem[Naur(1985)]{naur1985programming}
Naur, P.
\newblock Programming as theory building.
\newblock \emph{Microprocessing and Microprogramming}, 15\penalty0 (5):\penalty0 253--261, 1985.
\newblock ISSN 0165-6074.
\newblock \doi{https://doi.org/10.1016/0165-6074(85)90032-8}.
\newblock URL \url{https://www.sciencedirect.com/science/article/pii/0165607485900328}.

\bibitem[OpenAI et~al.(2024)OpenAI, Achiam, Adler, Agarwal, Ahmad, Akkaya, Aleman, Almeida, Altenschmidt, Altman, Anadkat, Avila, Babuschkin, Balaji, Balcom, Baltescu, Bao, Bavarian, Belgum, Bello, Berdine, Bernadett-Shapiro, Berner, Bogdonoff, Boiko, Boyd, Brakman, Brockman, Brooks, Brundage, Button, Cai, Campbell, Cann, Carey, Carlson, Carmichael, Chan, Chang, Chantzis, Chen, Chen, Chen, Chen, Chen, Chess, Cho, Chu, Chung, Cummings, Currier, Dai, Decareaux, Degry, Deutsch, Deville, Dhar, Dohan, Dowling, Dunning, Ecoffet, Eleti, Eloundou, Farhi, Fedus, Felix, Fishman, Forte, Fulford, Gao, Georges, Gibson, Goel, Gogineni, Goh, Gontijo-Lopes, Gordon, Grafstein, Gray, Greene, Gross, Gu, Guo, Hallacy, Han, Harris, He, Heaton, Heidecke, Hesse, Hickey, Hickey, Hoeschele, Houghton, Hsu, Hu, Hu, Huizinga, Jain, Jain, Jang, Jiang, Jiang, Jin, Jin, Jomoto, Jonn, Jun, Kaftan, Łukasz Kaiser, Kamali, Kanitscheider, Keskar, Khan, Kilpatrick, Kim, Kim, Kim, Kirchner, Kiros, Knight, Kokotajlo, Łukasz Kondraciuk, Kondrich,
  Konstantinidis, Kosic, Krueger, Kuo, Lampe, Lan, Lee, Leike, Leung, Levy, Li, Lim, Lin, Lin, Litwin, Lopez, Lowe, Lue, Makanju, Malfacini, Manning, Markov, Markovski, Martin, Mayer, Mayne, McGrew, McKinney, McLeavey, McMillan, McNeil, Medina, Mehta, Menick, Metz, Mishchenko, Mishkin, Monaco, Morikawa, Mossing, Mu, Murati, Murk, Mély, Nair, Nakano, Nayak, Neelakantan, Ngo, Noh, Ouyang, O'Keefe, Pachocki, Paino, Palermo, Pantuliano, Parascandolo, Parish, Parparita, Passos, Pavlov, Peng, Perelman, de~Avila Belbute~Peres, Petrov, de~Oliveira~Pinto, Michael, Pokorny, Pokrass, Pong, Powell, Power, Power, Proehl, Puri, Radford, Rae, Ramesh, Raymond, Real, Rimbach, Ross, Rotsted, Roussez, Ryder, Saltarelli, Sanders, Santurkar, Sastry, Schmidt, Schnurr, Schulman, Selsam, Sheppard, Sherbakov, Shieh, Shoker, Shyam, Sidor, Sigler, Simens, Sitkin, Slama, Sohl, Sokolowsky, Song, Staudacher, Such, Summers, Sutskever, Tang, Tezak, Thompson, Tillet, Tootoonchian, Tseng, Tuggle, Turley, Tworek, Uribe, Vallone, Vijayvergiya,
  Voss, Wainwright, Wang, Wang, Wang, Ward, Wei, Weinmann, Welihinda, Welinder, Weng, Weng, Wiethoff, Willner, Winter, Wolrich, Wong, Workman, Wu, Wu, Wu, Xiao, Xu, Yoo, Yu, Yuan, Zaremba, Zellers, Zhang, Zhang, Zhao, Zheng, Zhuang, Zhuk, and Zoph]{openai2024gpt4technicalreport}
OpenAI, Achiam, J., Adler, S., Agarwal, S., Ahmad, L., Akkaya, I., Aleman, F.~L., Almeida, D., Altenschmidt, J., Altman, S., Anadkat, S., Avila, R., Babuschkin, I., Balaji, S., Balcom, V., Baltescu, P., Bao, H., Bavarian, M., Belgum, J., Bello, I., Berdine, J., Bernadett-Shapiro, G., Berner, C., Bogdonoff, L., Boiko, O., Boyd, M., Brakman, A.-L., Brockman, G., Brooks, T., Brundage, M., Button, K., Cai, T., Campbell, R., Cann, A., Carey, B., Carlson, C., Carmichael, R., Chan, B., Chang, C., Chantzis, F., Chen, D., Chen, S., Chen, R., Chen, J., Chen, M., Chess, B., Cho, C., Chu, C., Chung, H.~W., Cummings, D., Currier, J., Dai, Y., Decareaux, C., Degry, T., Deutsch, N., Deville, D., Dhar, A., Dohan, D., Dowling, S., Dunning, S., Ecoffet, A., Eleti, A., Eloundou, T., Farhi, D., Fedus, L., Felix, N., Fishman, S.~P., Forte, J., Fulford, I., Gao, L., Georges, E., Gibson, C., Goel, V., Gogineni, T., Goh, G., Gontijo-Lopes, R., Gordon, J., Grafstein, M., Gray, S., Greene, R., Gross, J., Gu, S.~S., Guo, Y., Hallacy,
  C., Han, J., Harris, J., He, Y., Heaton, M., Heidecke, J., Hesse, C., Hickey, A., Hickey, W., Hoeschele, P., Houghton, B., Hsu, K., Hu, S., Hu, X., Huizinga, J., Jain, S., Jain, S., Jang, J., Jiang, A., Jiang, R., Jin, H., Jin, D., Jomoto, S., Jonn, B., Jun, H., Kaftan, T., Łukasz Kaiser, Kamali, A., Kanitscheider, I., Keskar, N.~S., Khan, T., Kilpatrick, L., Kim, J.~W., Kim, C., Kim, Y., Kirchner, J.~H., Kiros, J., Knight, M., Kokotajlo, D., Łukasz Kondraciuk, Kondrich, A., Konstantinidis, A., Kosic, K., Krueger, G., Kuo, V., Lampe, M., Lan, I., Lee, T., Leike, J., Leung, J., Levy, D., Li, C.~M., Lim, R., Lin, M., Lin, S., Litwin, M., Lopez, T., Lowe, R., Lue, P., Makanju, A., Malfacini, K., Manning, S., Markov, T., Markovski, Y., Martin, B., Mayer, K., Mayne, A., McGrew, B., McKinney, S.~M., McLeavey, C., McMillan, P., McNeil, J., Medina, D., Mehta, A., Menick, J., Metz, L., Mishchenko, A., Mishkin, P., Monaco, V., Morikawa, E., Mossing, D., Mu, T., Murati, M., Murk, O., Mély, D., Nair, A., Nakano, R.,
  Nayak, R., Neelakantan, A., Ngo, R., Noh, H., Ouyang, L., O'Keefe, C., Pachocki, J., Paino, A., Palermo, J., Pantuliano, A., Parascandolo, G., Parish, J., Parparita, E., Passos, A., Pavlov, M., Peng, A., Perelman, A., de~Avila Belbute~Peres, F., Petrov, M., de~Oliveira~Pinto, H.~P., Michael, Pokorny, Pokrass, M., Pong, V.~H., Powell, T., Power, A., Power, B., Proehl, E., Puri, R., Radford, A., Rae, J., Ramesh, A., Raymond, C., Real, F., Rimbach, K., Ross, C., Rotsted, B., Roussez, H., Ryder, N., Saltarelli, M., Sanders, T., Santurkar, S., Sastry, G., Schmidt, H., Schnurr, D., Schulman, J., Selsam, D., Sheppard, K., Sherbakov, T., Shieh, J., Shoker, S., Shyam, P., Sidor, S., Sigler, E., Simens, M., Sitkin, J., Slama, K., Sohl, I., Sokolowsky, B., Song, Y., Staudacher, N., Such, F.~P., Summers, N., Sutskever, I., Tang, J., Tezak, N., Thompson, M.~B., Tillet, P., Tootoonchian, A., Tseng, E., Tuggle, P., Turley, N., Tworek, J., Uribe, J. F.~C., Vallone, A., Vijayvergiya, A., Voss, C., Wainwright, C., Wang,
  J.~J., Wang, A., Wang, B., Ward, J., Wei, J., Weinmann, C., Welihinda, A., Welinder, P., Weng, J., Weng, L., Wiethoff, M., Willner, D., Winter, C., Wolrich, S., Wong, H., Workman, L., Wu, S., Wu, J., Wu, M., Xiao, K., Xu, T., Yoo, S., Yu, K., Yuan, Q., Zaremba, W., Zellers, R., Zhang, C., Zhang, M., Zhao, S., Zheng, T., Zhuang, J., Zhuk, W., and Zoph, B.
\newblock Gpt-4 technical report, 2024.
\newblock URL \url{https://arxiv.org/abs/2303.08774}.

\bibitem[Paglieri et~al.(2024)Paglieri, Cupia{\l}, Coward, Piterbarg, Wolczyk, Khan, Pignatelli, Kuci{\'n}ski, Pinto, Fergus, et~al.]{paglieri2024balrog}
Paglieri, D., Cupia{\l}, B., Coward, S., Piterbarg, U., Wolczyk, M., Khan, A., Pignatelli, E., Kuci{\'n}ski, {\L}., Pinto, L., Fergus, R., et~al.
\newblock Balrog: Benchmarking agentic llm and vlm reasoning on games.
\newblock \emph{arXiv preprint arXiv:2411.13543}, 2024.

\bibitem[Reid et~al.(2021)Reid, Miralavy, Kelly, Banzhaf, and Gondro]{reid2021factorygrowautomationfactorio}
Reid, K.~N., Miralavy, I., Kelly, S., Banzhaf, W., and Gondro, C.
\newblock The factory must grow: Automation in factorio, 2021.
\newblock URL \url{https://arxiv.org/abs/2102.04871}.

\bibitem[Ruan et~al.(2023)Ruan, Chen, Zhang, Xu, Bao, Mao, Li, Zeng, Zhao, et~al.]{ruan2023tptu}
Ruan, J., Chen, Y., Zhang, B., Xu, Z., Bao, T., Mao, H., Li, Z., Zeng, X., Zhao, R., et~al.
\newblock Tptu: Task planning and tool usage of large language model-based ai agents.
\newblock In \emph{NeurIPS 2023 Foundation Models for Decision Making Workshop}, 2023.

\bibitem[Shridhar et~al.(2020)Shridhar, Mottaghi, Kolve, and Gupta]{alfworld}
Shridhar, M., Mottaghi, R., Kolve, E., and Gupta, A.
\newblock {ALFWorld: Aligning Text and Embodied Environments for Interactive Learning}.
\newblock \url{https://github.com/alfworld/alfworld}, 2020.
\newblock Accessed: 2025-01-07.

\bibitem[Silver et~al.(2016)Silver, Huang, Maddison, and et~al.]{AlphaGo}
Silver, D., Huang, A., Maddison, C.~J., and et~al.
\newblock {Mastering the game of Go with deep neural networks and tree search}.
\newblock \emph{Nature}, 529\penalty0 (7587):\penalty0 484--489, 2016.
\newblock \doi{10.1038/nature16961}.

\bibitem[Skalse et~al.(2022)Skalse, Howe, Krasheninnikov, and Krueger]{skalse2022defining}
Skalse, J., Howe, N., Krasheninnikov, D., and Krueger, D.
\newblock Defining and characterizing reward gaming.
\newblock \emph{Advances in Neural Information Processing Systems}, 35:\penalty0 9460--9471, 2022.

\bibitem[Team et~al.(2024)Team, Anil, Borgeaud, Alayrac, Yu, Soricut, Schalkwyk, Dai, Hauth, Millican, Silver, Johnson, Antonoglou, Schrittwieser, Glaese, Chen, Pitler, Lillicrap, Lazaridou, Firat, Molloy, Isard, Barham, Hennigan, Lee, Viola, Reynolds, Xu, Doherty, Collins, Meyer, Rutherford, Moreira, Ayoub, Goel, Krawczyk, Du, Chi, Cheng, Ni, Shah, Kane, Chan, Faruqui, Severyn, Lin, Li, Cheng, Ittycheriah, Mahdieh, Chen, Sun, Tran, Bagri, Lakshminarayanan, Liu, Orban, Güra, Zhou, Song, Boffy, Ganapathy, Zheng, Choe, Ágoston Weisz, Zhu, Lu, Gopal, Kahn, Kula, Pitman, Shah, Taropa, Merey, Baeuml, Chen, Shafey, Zhang, Sercinoglu, Tucker, Piqueras, Krikun, Barr, Savinov, Danihelka, Roelofs, White, Andreassen, von Glehn, Yagati, Kazemi, Gonzalez, Khalman, Sygnowski, Frechette, Smith, Culp, Proleev, Luan, Chen, Lottes, Schucher, Lebron, Rrustemi, Clay, Crone, Kocisky, Zhao, Perz, Yu, Howard, Bloniarz, Rae, Lu, Sifre, Maggioni, Alcober, Garrette, Barnes, Thakoor, Austin, Barth-Maron, Wong, Joshi, Chaabouni,
  Fatiha, Ahuja, Tomar, Senter, Chadwick, Kornakov, Attaluri, Iturrate, Liu, Li, Cogan, Chen, Jia, Gu, Zhang, Grimstad, Hartman, Garcia, Pillai, Devlin, Laskin, de~Las~Casas, Valter, Tao, Blanco, Badia, Reitter, Chen, Brennan, Rivera, Brin, Iqbal, Surita, Labanowski, Rao, Winkler, Parisotto, Gu, Olszewska, Addanki, Miech, Louis, Teplyashin, Brown, Catt, Balaguer, Xiang, Wang, Ashwood, Briukhov, Webson, Ganapathy, Sanghavi, Kannan, Chang, Stjerngren, Djolonga, Sun, Bapna, Aitchison, Pejman, Michalewski, Yu, Wang, Love, Ahn, Bloxwich, Han, Humphreys, Sellam, Bradbury, Godbole, Samangooei, Damoc, Kaskasoli, Arnold, Vasudevan, Agrawal, Riesa, Lepikhin, Tanburn, Srinivasan, Lim, Hodkinson, Shyam, Ferret, Hand, Garg, Paine, Li, Li, Giang, Neitz, Abbas, York, Reid, Cole, Chowdhery, Das, Rogozińska, Nikolaev, Sprechmann, Nado, Zilka, Prost, He, Monteiro, Mishra, Welty, Newlan, Jia, Allamanis, Hu, de~Liedekerke, Gilmer, Saroufim, Rijhwani, Hou, Shrivastava, Baddepudi, Goldin, Ozturel, Cassirer, Xu, Sohn, Sachan,
  Amplayo, Swanson, Petrova, Narayan, Guez, Brahma, Landon, Patel, Zhao, Villela, Wang, Jia, Rahtz, Giménez, Yeung, Keeling, Georgiev, Mincu, Wu, Haykal, Saputro, Vodrahalli, Qin, Cankara, Sharma, Fernando, Hawkins, Neyshabur, Kim, Hutter, Agrawal, Castro-Ros, van~den Driessche, Wang, Yang, yiin Chang, Komarek, McIlroy, Lučić, Zhang, Farhan, Sharman, Natsev, Michel, Bansal, Qiao, Cao, Shakeri, Butterfield, Chung, Rubenstein, Agrawal, Mensch, Soparkar, Lenc, Chung, Pope, Maggiore, Kay, Jhakra, Wang, Maynez, Phuong, Tobin, Tacchetti, Trebacz, Robinson, Katariya, Riedel, Bailey, Xiao, Ghelani, Aroyo, Slone, Houlsby, Xiong, Yang, Gribovskaya, Adler, Wirth, Lee, Li, Kagohara, Pavagadhi, Bridgers, Bortsova, Ghemawat, Ahmed, Liu, Powell, Bolina, Iinuma, Zablotskaia, Besley, Chung, Dozat, Comanescu, Si, Greer, Su, Polacek, Kaufman, Tokumine, Hu, Buchatskaya, Miao, Elhawaty, Siddhant, Tomasev, Xing, Greer, Miller, Ashraf, Roy, Zhang, Ma, Filos, Besta, Blevins, Klimenko, Yeh, Changpinyo, Mu, Chang, Pajarskas, Muir,
  Cohen, Lan, Haridasan, Marathe, Hansen, Douglas, Samuel, Wang, Austin, Lan, Jiang, Chiu, Lorenzo, Sjösund, Cevey, Gleicher, Avrahami, Boral, Srinivasan, Selo, May, Aisopos, Hussenot, Soares, Baumli, Chang, Recasens, Caine, Pritzel, Pavetic, Pardo, Gergely, Frye, Ramasesh, Horgan, Badola, Kassner, Roy, Dyer, Campos, Tomala, Tang, Badawy, White, Mustafa, Lang, Jindal, Vikram, Gong, Caelles, Hemsley, Thornton, Feng, Stokowiec, Zheng, Thacker, Çağlar Ünlü, Zhang, Saleh, Svensson, Bileschi, Patil, Anand, Ring, Tsihlas, Vezer, Selvi, Shevlane, Rodriguez, Kwiatkowski, Daruki, Rong, Dafoe, FitzGerald, Gu-Lemberg, Khan, Hendricks, Pellat, Feinberg, Cobon-Kerr, Sainath, Rauh, Hashemi, Ives, Hasson, Noland, Cao, Byrd, Hou, Wang, Sottiaux, Paganini, Lespiau, Moufarek, Hassan, Shivakumar, van Amersfoort, Mandhane, Joshi, Goyal, Tung, Brock, Sheahan, Misra, Li, Rakićević, Dehghani, Liu, Mittal, Oh, Noury, Sezener, Huot, Lamm, Cao, Chen, Mudgal, Stella, Brooks, Vasudevan, Liu, Chain, Melinkeri, Cohen, Wang,
  Seymore, Zubkov, Goel, Yue, Krishnakumaran, Albert, Hurley, Sano, Mohananey, Joughin, Filonov, Kepa, Eldawy, Lim, Rishi, Badiezadegan, Bos, Chang, Jain, Padmanabhan, Puttagunta, Krishna, Baker, Kalb, Bedapudi, Kurzrok, Lei, Yu, Litvin, Zhou, Wu, Sobell, Siciliano, Papir, Neale, Bragagnolo, Toor, Chen, Anklin, Wang, Feng, Gholami, Ling, Liu, Walter, Moghaddam, Kishore, Adamek, Mercado, Mallinson, Wandekar, Cagle, Ofek, Garrido, Lombriser, Mukha, Sun, Mohammad, Matak, Qian, Peswani, Janus, Yuan, Schelin, David, Garg, He, Duzhyi, Älgmyr, Lottaz, Li, Yadav, Xu, Chinien, Shivanna, Chuklin, Li, Spadine, Wolfe, Mohamed, Das, Dai, He, von Dincklage, Upadhyay, Maurya, Chi, Krause, Salama, Rabinovitch, M, Selvan, Dektiarev, Ghiasi, Guven, Gupta, Liu, Sharma, Shtacher, Paul, Akerlund, Aubet, Huang, Zhu, Zhu, Teixeira, Fritze, Bertolini, Marinescu, Bölle, Paulus, Gupta, Latkar, Chang, Sanders, Wilson, Wu, Tan, Thiet, Doshi, Lall, Mishra, Chen, Luong, Benjamin, Lee, Andrejczuk, Rabiej, Ranjan, Styrc, Yin, Simon,
  Harriott, Bansal, Robsky, Bacon, Greene, Mirylenka, Zhou, Sarvana, Goyal, Andermatt, Siegler, Horn, Israel, Pongetti, Chen, Selvatici, Silva, Wang, Tolins, Guu, Yogev, Cai, Agostini, Shah, Nguyen, Donnaile, Pereira, Friso, Stambler, Kurzrok, Kuang, Romanikhin, Geller, Yan, Jang, Lee, Fica, Malmi, Tan, Banica, Balle, Pham, Huang, Avram, Shi, Singh, Hidey, Ahuja, Saxena, Dooley, Potharaju, O'Neill, Gokulchandran, Foley, Zhao, Dusenberry, Liu, Mehta, Kotikalapudi, Safranek-Shrader, Goodman, Kessinger, Globen, Kolhar, Gorgolewski, Ibrahim, Song, Eichenbaum, Brovelli, Potluri, Lahoti, Baetu, Ghorbani, Chen, Crawford, Pal, Sridhar, Gurita, Mujika, Petrovski, Cedoz, Li, Chen, Santo, Goyal, Punjabi, Kappaganthu, Kwak, LV, Velury, Choudhury, Hall, Shah, Figueira, Thomas, Lu, Zhou, Kumar, Jurdi, Chikkerur, Ma, Yu, Kwak, Ähdel, Rajayogam, Choma, Liu, Barua, Ji, Park, Hellendoorn, Bailey, Bilal, Zhou, Khatir, Sutton, Rzadkowski, Macintosh, Shagin, Medina, Liang, Zhou, Shah, Bi, Dankovics, Banga, Lehmann, Bredesen,
  Lin, Hoffmann, Lai, Chung, Yang, Balani, Bražinskas, Sozanschi, Hayes, Alcalde, Makarov, Chen, Stella, Snijders, Mandl, Kärrman, Nowak, Wu, Dyck, Vaidyanathan, R, Mallet, Rudominer, Johnston, Mittal, Udathu, Christensen, Verma, Irving, Santucci, Elsayed, Davoodi, Georgiev, Tenney, Hua, Cideron, Leurent, Alnahlawi, Georgescu, Wei, Zheng, Scandinaro, Jiang, Snoek, Sundararajan, Wang, Ontiveros, Karo, Cole, Rajashekhar, Tumeh, Ben-David, Jain, Uesato, Datta, Bunyan, Wu, Zhang, Stanczyk, Zhang, Steiner, Naskar, Azzam, Johnson, Paszke, Chiu, Elias, Mohiuddin, Muhammad, Miao, Lee, Vieillard, Park, Zhang, Stanway, Garmon, Karmarkar, Dong, Lee, Kumar, Zhou, Evens, Isaac, Irving, Loper, Fink, Arkatkar, Chen, Shafran, Petrychenko, Chen, Jia, Levskaya, Zhu, Grabowski, Mao, Magni, Yao, Snaider, Casagrande, Palmer, Suganthan, Castaño, Giannoumis, Kim, Rybiński, Sreevatsa, Prendki, Soergel, Goedeckemeyer, Gierke, Jafari, Gaba, Wiesner, Wright, Wei, Vashisht, Kulizhskaya, Hoover, Le, Li, Iwuanyanwu, Liu, Ramirez,
  Khorlin, Cui, LIN, Wu, Aguilar, Pallo, Chakladar, Perng, Abellan, Zhang, Dasgupta, Kushman, Penchev, Repina, Wu, van~der Weide, Ponnapalli, Kaplan, Simsa, Li, Dousse, Yang, Piper, Ie, Pasumarthi, Lintz, Vijayakumar, Andor, Valenzuela, Lui, Paduraru, Peng, Lee, Zhang, Greene, Nguyen, Kurylowicz, Hardin, Dixon, Janzer, Choo, Feng, Zhang, Singhal, Du, McKinnon, Antropova, Bolukbasi, Keller, Reid, Finchelstein, Raad, Crocker, Hawkins, Dadashi, Gaffney, Franko, Bulanova, Leblond, Chung, Askham, Cobo, Xu, Fischer, Xu, Sorokin, Alberti, Lin, Evans, Dimitriev, Forbes, Banarse, Tung, Omernick, Bishop, Sterneck, Jain, Xia, Amid, Piccinno, Wang, Banzal, Mankowitz, Polozov, Krakovna, Brown, Bateni, Duan, Firoiu, Thotakuri, Natan, Geist, tan Girgin, Li, Ye, Roval, Tojo, Kwong, Lee-Thorp, Yew, Sinopalnikov, Ramos, Mellor, Sharma, Wu, Miller, Sonnerat, Vnukov, Greig, Beattie, Caveness, Bai, Eisenschlos, Korchemniy, Tsai, Jasarevic, Kong, Dao, Zheng, Liu, Yang, Zhu, Teh, Sanmiya, Gladchenko, Trdin, Toyama, Rosen, Tavakkol,
  Xue, Elkind, Woodman, Carpenter, Papamakarios, Kemp, Kafle, Grunina, Sinha, Talbert, Wu, Owusu-Afriyie, Du, Thornton, Pont-Tuset, Narayana, Li, Fatehi, Wieting, Ajmeri, Uria, Ko, Knight, Héliou, Niu, Gu, Pang, Li, Levine, Stolovich, Santamaria-Fernandez, Goenka, Yustalim, Strudel, Elqursh, Deck, Lee, Li, Levin, Hoffmann, Holtmann-Rice, Bachem, Arora, Koh, Yeganeh, Põder, Tariq, Sun, Ionita, Seyedhosseini, Tafti, Liu, Gulati, Liu, Ye, Chrzaszcz, Wang, Sethi, Li, Brown, Singh, Fan, Parisi, Stanton, Koverkathu, Choquette-Choo, Li, Lu, Ittycheriah, Shroff, Varadarajan, Bahargam, Willoughby, Gaddy, Desjardins, Cornero, Robenek, Mittal, Albrecht, Shenoy, Moiseev, Jacobsson, Ghaffarkhah, Rivière, Walton, Crepy, Parrish, Zhou, Farabet, Radebaugh, Srinivasan, van~der Salm, Fidjeland, Scellato, Latorre-Chimoto, Klimczak-Plucińska, Bridson, de~Cesare, Hudson, Mendolicchio, Walker, Morris, Mauger, Guseynov, Reid, Odoom, Loher, Cotruta, Yenugula, Grewe, Petrushkina, Duerig, Sanchez, Yadlowsky, Shen, Globerson, Webb,
  Dua, Li, Bhupatiraju, Hurt, Qureshi, Agarwal, Shani, Eyal, Khare, Belle, Wang, Tekur, Kale, Wei, Sang, Saeta, Liechty, Sun, Zhao, Lee, Nayak, Fritz, Vuyyuru, Aslanides, Vyas, Wicke, Ma, Eltyshev, Martin, Cate, Manyika, Amiri, Kim, Xiong, Kang, Luisier, Tripuraneni, Madras, Guo, Waters, Wang, Ainslie, Baldridge, Zhang, Pruthi, Bauer, Yang, Mansour, Gelman, Xu, Polovets, Liu, Cai, Chen, Sheng, Xue, Ozair, Angermueller, Li, Sinha, Wang, Wiesinger, Koukoumidis, Tian, Iyer, Gurumurthy, Goldenson, Shah, Blake, Yu, Urbanowicz, Palomaki, Fernando, Durden, Mehta, Momchev, Rahimtoroghi, Georgaki, Raul, Ruder, Redshaw, Lee, Zhou, Jalan, Li, Hechtman, Schuh, Nasr, Milan, Mikulik, Franco, Green, Nguyen, Kelley, Mahendru, Hu, Howland, Vargas, Hui, Bansal, Rao, Ghiya, Wang, Ye, Sarr, Preston, Elish, Li, Kaku, Gupta, Pasupat, Juan, Someswar, M., Chen, Amini, Fabrikant, Chu, Dong, Muthal, Buthpitiya, Jauhari, Hua, Khandelwal, Hitron, Ren, Rinaldi, Drath, Dabush, Jiang, Godhia, Sachs, Chen, Fan, Taitelbaum, Noga, Dai, Wang,
  Liang, Hamer, Ferng, Elkind, Atias, Lee, Listík, Carlen, van~de Kerkhof, Pikus, Zaher, Müller, Zykova, Stefanec, Gatsko, Hirnschall, Sethi, Xu, Ahuja, Tsai, Stefanoiu, Feng, Dhandhania, Katyal, Gupta, Parulekar, Pitta, Zhao, Bhatia, Bhavnani, Alhadlaq, Li, Danenberg, Tu, Pine, Filippova, Ghosh, Limonchik, Urala, Lanka, Clive, Sun, Li, Wu, Hongtongsak, Li, Thakkar, Omarov, Majmundar, Alverson, Kucharski, Patel, Jain, Zabelin, Pelagatti, Kohli, Kumar, Kim, Sankar, Shah, Ramachandruni, Zeng, Bariach, Weidinger, Vu, Andreev, He, Hui, Kashem, Subramanya, Hsiao, Hassabis, Kavukcuoglu, Sadovsky, Le, Strohman, Wu, Petrov, Dean, and Vinyals]{geminiteam2024geminifamilyhighlycapable}
Team, G., Anil, R., Borgeaud, S., Alayrac, J.-B., Yu, J., Soricut, R., Schalkwyk, J., Dai, A.~M., Hauth, A., Millican, K., Silver, D., Johnson, M., Antonoglou, I., Schrittwieser, J., Glaese, A., Chen, J., Pitler, E., Lillicrap, T., Lazaridou, A., Firat, O., Molloy, J., Isard, M., Barham, P.~R., Hennigan, T., Lee, B., Viola, F., Reynolds, M., Xu, Y., Doherty, R., Collins, E., Meyer, C., Rutherford, E., Moreira, E., Ayoub, K., Goel, M., Krawczyk, J., Du, C., Chi, E., Cheng, H.-T., Ni, E., Shah, P., Kane, P., Chan, B., Faruqui, M., Severyn, A., Lin, H., Li, Y., Cheng, Y., Ittycheriah, A., Mahdieh, M., Chen, M., Sun, P., Tran, D., Bagri, S., Lakshminarayanan, B., Liu, J., Orban, A., Güra, F., Zhou, H., Song, X., Boffy, A., Ganapathy, H., Zheng, S., Choe, H., Ágoston Weisz, Zhu, T., Lu, Y., Gopal, S., Kahn, J., Kula, M., Pitman, J., Shah, R., Taropa, E., Merey, M.~A., Baeuml, M., Chen, Z., Shafey, L.~E., Zhang, Y., Sercinoglu, O., Tucker, G., Piqueras, E., Krikun, M., Barr, I., Savinov, N., Danihelka, I.,
  Roelofs, B., White, A., Andreassen, A., von Glehn, T., Yagati, L., Kazemi, M., Gonzalez, L., Khalman, M., Sygnowski, J., Frechette, A., Smith, C., Culp, L., Proleev, L., Luan, Y., Chen, X., Lottes, J., Schucher, N., Lebron, F., Rrustemi, A., Clay, N., Crone, P., Kocisky, T., Zhao, J., Perz, B., Yu, D., Howard, H., Bloniarz, A., Rae, J.~W., Lu, H., Sifre, L., Maggioni, M., Alcober, F., Garrette, D., Barnes, M., Thakoor, S., Austin, J., Barth-Maron, G., Wong, W., Joshi, R., Chaabouni, R., Fatiha, D., Ahuja, A., Tomar, G.~S., Senter, E., Chadwick, M., Kornakov, I., Attaluri, N., Iturrate, I., Liu, R., Li, Y., Cogan, S., Chen, J., Jia, C., Gu, C., Zhang, Q., Grimstad, J., Hartman, A.~J., Garcia, X., Pillai, T.~S., Devlin, J., Laskin, M., de~Las~Casas, D., Valter, D., Tao, C., Blanco, L., Badia, A.~P., Reitter, D., Chen, M., Brennan, J., Rivera, C., Brin, S., Iqbal, S., Surita, G., Labanowski, J., Rao, A., Winkler, S., Parisotto, E., Gu, Y., Olszewska, K., Addanki, R., Miech, A., Louis, A., Teplyashin, D.,
  Brown, G., Catt, E., Balaguer, J., Xiang, J., Wang, P., Ashwood, Z., Briukhov, A., Webson, A., Ganapathy, S., Sanghavi, S., Kannan, A., Chang, M.-W., Stjerngren, A., Djolonga, J., Sun, Y., Bapna, A., Aitchison, M., Pejman, P., Michalewski, H., Yu, T., Wang, C., Love, J., Ahn, J., Bloxwich, D., Han, K., Humphreys, P., Sellam, T., Bradbury, J., Godbole, V., Samangooei, S., Damoc, B., Kaskasoli, A., Arnold, S. M.~R., Vasudevan, V., Agrawal, S., Riesa, J., Lepikhin, D., Tanburn, R., Srinivasan, S., Lim, H., Hodkinson, S., Shyam, P., Ferret, J., Hand, S., Garg, A., Paine, T.~L., Li, J., Li, Y., Giang, M., Neitz, A., Abbas, Z., York, S., Reid, M., Cole, E., Chowdhery, A., Das, D., Rogozińska, D., Nikolaev, V., Sprechmann, P., Nado, Z., Zilka, L., Prost, F., He, L., Monteiro, M., Mishra, G., Welty, C., Newlan, J., Jia, D., Allamanis, M., Hu, C.~H., de~Liedekerke, R., Gilmer, J., Saroufim, C., Rijhwani, S., Hou, S., Shrivastava, D., Baddepudi, A., Goldin, A., Ozturel, A., Cassirer, A., Xu, Y., Sohn, D., Sachan,
  D., Amplayo, R.~K., Swanson, C., Petrova, D., Narayan, S., Guez, A., Brahma, S., Landon, J., Patel, M., Zhao, R., Villela, K., Wang, L., Jia, W., Rahtz, M., Giménez, M., Yeung, L., Keeling, J., Georgiev, P., Mincu, D., Wu, B., Haykal, S., Saputro, R., Vodrahalli, K., Qin, J., Cankara, Z., Sharma, A., Fernando, N., Hawkins, W., Neyshabur, B., Kim, S., Hutter, A., Agrawal, P., Castro-Ros, A., van~den Driessche, G., Wang, T., Yang, F., yiin Chang, S., Komarek, P., McIlroy, R., Lučić, M., Zhang, G., Farhan, W., Sharman, M., Natsev, P., Michel, P., Bansal, Y., Qiao, S., Cao, K., Shakeri, S., Butterfield, C., Chung, J., Rubenstein, P.~K., Agrawal, S., Mensch, A., Soparkar, K., Lenc, K., Chung, T., Pope, A., Maggiore, L., Kay, J., Jhakra, P., Wang, S., Maynez, J., Phuong, M., Tobin, T., Tacchetti, A., Trebacz, M., Robinson, K., Katariya, Y., Riedel, S., Bailey, P., Xiao, K., Ghelani, N., Aroyo, L., Slone, A., Houlsby, N., Xiong, X., Yang, Z., Gribovskaya, E., Adler, J., Wirth, M., Lee, L., Li, M., Kagohara, T.,
  Pavagadhi, J., Bridgers, S., Bortsova, A., Ghemawat, S., Ahmed, Z., Liu, T., Powell, R., Bolina, V., Iinuma, M., Zablotskaia, P., Besley, J., Chung, D.-W., Dozat, T., Comanescu, R., Si, X., Greer, J., Su, G., Polacek, M., Kaufman, R.~L., Tokumine, S., Hu, H., Buchatskaya, E., Miao, Y., Elhawaty, M., Siddhant, A., Tomasev, N., Xing, J., Greer, C., Miller, H., Ashraf, S., Roy, A., Zhang, Z., Ma, A., Filos, A., Besta, M., Blevins, R., Klimenko, T., Yeh, C.-K., Changpinyo, S., Mu, J., Chang, O., Pajarskas, M., Muir, C., Cohen, V., Lan, C.~L., Haridasan, K., Marathe, A., Hansen, S., Douglas, S., Samuel, R., Wang, M., Austin, S., Lan, C., Jiang, J., Chiu, J., Lorenzo, J.~A., Sjösund, L.~L., Cevey, S., Gleicher, Z., Avrahami, T., Boral, A., Srinivasan, H., Selo, V., May, R., Aisopos, K., Hussenot, L., Soares, L.~B., Baumli, K., Chang, M.~B., Recasens, A., Caine, B., Pritzel, A., Pavetic, F., Pardo, F., Gergely, A., Frye, J., Ramasesh, V., Horgan, D., Badola, K., Kassner, N., Roy, S., Dyer, E., Campos, V.~C.,
  Tomala, A., Tang, Y., Badawy, D.~E., White, E., Mustafa, B., Lang, O., Jindal, A., Vikram, S., Gong, Z., Caelles, S., Hemsley, R., Thornton, G., Feng, F., Stokowiec, W., Zheng, C., Thacker, P., Çağlar Ünlü, Zhang, Z., Saleh, M., Svensson, J., Bileschi, M., Patil, P., Anand, A., Ring, R., Tsihlas, K., Vezer, A., Selvi, M., Shevlane, T., Rodriguez, M., Kwiatkowski, T., Daruki, S., Rong, K., Dafoe, A., FitzGerald, N., Gu-Lemberg, K., Khan, M., Hendricks, L.~A., Pellat, M., Feinberg, V., Cobon-Kerr, J., Sainath, T., Rauh, M., Hashemi, S.~H., Ives, R., Hasson, Y., Noland, E., Cao, Y., Byrd, N., Hou, L., Wang, Q., Sottiaux, T., Paganini, M., Lespiau, J.-B., Moufarek, A., Hassan, S., Shivakumar, K., van Amersfoort, J., Mandhane, A., Joshi, P., Goyal, A., Tung, M., Brock, A., Sheahan, H., Misra, V., Li, C., Rakićević, N., Dehghani, M., Liu, F., Mittal, S., Oh, J., Noury, S., Sezener, E., Huot, F., Lamm, M., Cao, N.~D., Chen, C., Mudgal, S., Stella, R., Brooks, K., Vasudevan, G., Liu, C., Chain, M., Melinkeri,
  N., Cohen, A., Wang, V., Seymore, K., Zubkov, S., Goel, R., Yue, S., Krishnakumaran, S., Albert, B., Hurley, N., Sano, M., Mohananey, A., Joughin, J., Filonov, E., Kepa, T., Eldawy, Y., Lim, J., Rishi, R., Badiezadegan, S., Bos, T., Chang, J., Jain, S., Padmanabhan, S. G.~S., Puttagunta, S., Krishna, K., Baker, L., Kalb, N., Bedapudi, V., Kurzrok, A., Lei, S., Yu, A., Litvin, O., Zhou, X., Wu, Z., Sobell, S., Siciliano, A., Papir, A., Neale, R., Bragagnolo, J., Toor, T., Chen, T., Anklin, V., Wang, F., Feng, R., Gholami, M., Ling, K., Liu, L., Walter, J., Moghaddam, H., Kishore, A., Adamek, J., Mercado, T., Mallinson, J., Wandekar, S., Cagle, S., Ofek, E., Garrido, G., Lombriser, C., Mukha, M., Sun, B., Mohammad, H.~R., Matak, J., Qian, Y., Peswani, V., Janus, P., Yuan, Q., Schelin, L., David, O., Garg, A., He, Y., Duzhyi, O., Älgmyr, A., Lottaz, T., Li, Q., Yadav, V., Xu, L., Chinien, A., Shivanna, R., Chuklin, A., Li, J., Spadine, C., Wolfe, T., Mohamed, K., Das, S., Dai, Z., He, K., von Dincklage, D.,
  Upadhyay, S., Maurya, A., Chi, L., Krause, S., Salama, K., Rabinovitch, P.~G., M, P. K.~R., Selvan, A., Dektiarev, M., Ghiasi, G., Guven, E., Gupta, H., Liu, B., Sharma, D., Shtacher, I.~H., Paul, S., Akerlund, O., Aubet, F.-X., Huang, T., Zhu, C., Zhu, E., Teixeira, E., Fritze, M., Bertolini, F., Marinescu, L.-E., Bölle, M., Paulus, D., Gupta, K., Latkar, T., Chang, M., Sanders, J., Wilson, R., Wu, X., Tan, Y.-X., Thiet, L.~N., Doshi, T., Lall, S., Mishra, S., Chen, W., Luong, T., Benjamin, S., Lee, J., Andrejczuk, E., Rabiej, D., Ranjan, V., Styrc, K., Yin, P., Simon, J., Harriott, M.~R., Bansal, M., Robsky, A., Bacon, G., Greene, D., Mirylenka, D., Zhou, C., Sarvana, O., Goyal, A., Andermatt, S., Siegler, P., Horn, B., Israel, A., Pongetti, F., Chen, C.-W.~L., Selvatici, M., Silva, P., Wang, K., Tolins, J., Guu, K., Yogev, R., Cai, X., Agostini, A., Shah, M., Nguyen, H., Donnaile, N.~O., Pereira, S., Friso, L., Stambler, A., Kurzrok, A., Kuang, C., Romanikhin, Y., Geller, M., Yan, Z., Jang, K., Lee,
  C.-C., Fica, W., Malmi, E., Tan, Q., Banica, D., Balle, D., Pham, R., Huang, Y., Avram, D., Shi, H., Singh, J., Hidey, C., Ahuja, N., Saxena, P., Dooley, D., Potharaju, S.~P., O'Neill, E., Gokulchandran, A., Foley, R., Zhao, K., Dusenberry, M., Liu, Y., Mehta, P., Kotikalapudi, R., Safranek-Shrader, C., Goodman, A., Kessinger, J., Globen, E., Kolhar, P., Gorgolewski, C., Ibrahim, A., Song, Y., Eichenbaum, A., Brovelli, T., Potluri, S., Lahoti, P., Baetu, C., Ghorbani, A., Chen, C., Crawford, A., Pal, S., Sridhar, M., Gurita, P., Mujika, A., Petrovski, I., Cedoz, P.-L., Li, C., Chen, S., Santo, N.~D., Goyal, S., Punjabi, J., Kappaganthu, K., Kwak, C., LV, P., Velury, S., Choudhury, H., Hall, J., Shah, P., Figueira, R., Thomas, M., Lu, M., Zhou, T., Kumar, C., Jurdi, T., Chikkerur, S., Ma, Y., Yu, A., Kwak, S., Ähdel, V., Rajayogam, S., Choma, T., Liu, F., Barua, A., Ji, C., Park, J.~H., Hellendoorn, V., Bailey, A., Bilal, T., Zhou, H., Khatir, M., Sutton, C., Rzadkowski, W., Macintosh, F., Shagin, K.,
  Medina, P., Liang, C., Zhou, J., Shah, P., Bi, Y., Dankovics, A., Banga, S., Lehmann, S., Bredesen, M., Lin, Z., Hoffmann, J.~E., Lai, J., Chung, R., Yang, K., Balani, N., Bražinskas, A., Sozanschi, A., Hayes, M., Alcalde, H.~F., Makarov, P., Chen, W., Stella, A., Snijders, L., Mandl, M., Kärrman, A., Nowak, P., Wu, X., Dyck, A., Vaidyanathan, K., R, R., Mallet, J., Rudominer, M., Johnston, E., Mittal, S., Udathu, A., Christensen, J., Verma, V., Irving, Z., Santucci, A., Elsayed, G., Davoodi, E., Georgiev, M., Tenney, I., Hua, N., Cideron, G., Leurent, E., Alnahlawi, M., Georgescu, I., Wei, N., Zheng, I., Scandinaro, D., Jiang, H., Snoek, J., Sundararajan, M., Wang, X., Ontiveros, Z., Karo, I., Cole, J., Rajashekhar, V., Tumeh, L., Ben-David, E., Jain, R., Uesato, J., Datta, R., Bunyan, O., Wu, S., Zhang, J., Stanczyk, P., Zhang, Y., Steiner, D., Naskar, S., Azzam, M., Johnson, M., Paszke, A., Chiu, C.-C., Elias, J.~S., Mohiuddin, A., Muhammad, F., Miao, J., Lee, A., Vieillard, N., Park, J., Zhang, J.,
  Stanway, J., Garmon, D., Karmarkar, A., Dong, Z., Lee, J., Kumar, A., Zhou, L., Evens, J., Isaac, W., Irving, G., Loper, E., Fink, M., Arkatkar, I., Chen, N., Shafran, I., Petrychenko, I., Chen, Z., Jia, J., Levskaya, A., Zhu, Z., Grabowski, P., Mao, Y., Magni, A., Yao, K., Snaider, J., Casagrande, N., Palmer, E., Suganthan, P., Castaño, A., Giannoumis, I., Kim, W., Rybiński, M., Sreevatsa, A., Prendki, J., Soergel, D., Goedeckemeyer, A., Gierke, W., Jafari, M., Gaba, M., Wiesner, J., Wright, D.~G., Wei, Y., Vashisht, H., Kulizhskaya, Y., Hoover, J., Le, M., Li, L., Iwuanyanwu, C., Liu, L., Ramirez, K., Khorlin, A., Cui, A., LIN, T., Wu, M., Aguilar, R., Pallo, K., Chakladar, A., Perng, G., Abellan, E.~A., Zhang, M., Dasgupta, I., Kushman, N., Penchev, I., Repina, A., Wu, X., van~der Weide, T., Ponnapalli, P., Kaplan, C., Simsa, J., Li, S., Dousse, O., Yang, F., Piper, J., Ie, N., Pasumarthi, R., Lintz, N., Vijayakumar, A., Andor, D., Valenzuela, P., Lui, M., Paduraru, C., Peng, D., Lee, K., Zhang, S.,
  Greene, S., Nguyen, D.~D., Kurylowicz, P., Hardin, C., Dixon, L., Janzer, L., Choo, K., Feng, Z., Zhang, B., Singhal, A., Du, D., McKinnon, D., Antropova, N., Bolukbasi, T., Keller, O., Reid, D., Finchelstein, D., Raad, M.~A., Crocker, R., Hawkins, P., Dadashi, R., Gaffney, C., Franko, K., Bulanova, A., Leblond, R., Chung, S., Askham, H., Cobo, L.~C., Xu, K., Fischer, F., Xu, J., Sorokin, C., Alberti, C., Lin, C.-C., Evans, C., Dimitriev, A., Forbes, H., Banarse, D., Tung, Z., Omernick, M., Bishop, C., Sterneck, R., Jain, R., Xia, J., Amid, E., Piccinno, F., Wang, X., Banzal, P., Mankowitz, D.~J., Polozov, A., Krakovna, V., Brown, S., Bateni, M., Duan, D., Firoiu, V., Thotakuri, M., Natan, T., Geist, M., tan Girgin, S., Li, H., Ye, J., Roval, O., Tojo, R., Kwong, M., Lee-Thorp, J., Yew, C., Sinopalnikov, D., Ramos, S., Mellor, J., Sharma, A., Wu, K., Miller, D., Sonnerat, N., Vnukov, D., Greig, R., Beattie, J., Caveness, E., Bai, L., Eisenschlos, J., Korchemniy, A., Tsai, T., Jasarevic, M., Kong, W., Dao,
  P., Zheng, Z., Liu, F., Yang, F., Zhu, R., Teh, T.~H., Sanmiya, J., Gladchenko, E., Trdin, N., Toyama, D., Rosen, E., Tavakkol, S., Xue, L., Elkind, C., Woodman, O., Carpenter, J., Papamakarios, G., Kemp, R., Kafle, S., Grunina, T., Sinha, R., Talbert, A., Wu, D., Owusu-Afriyie, D., Du, C., Thornton, C., Pont-Tuset, J., Narayana, P., Li, J., Fatehi, S., Wieting, J., Ajmeri, O., Uria, B., Ko, Y., Knight, L., Héliou, A., Niu, N., Gu, S., Pang, C., Li, Y., Levine, N., Stolovich, A., Santamaria-Fernandez, R., Goenka, S., Yustalim, W., Strudel, R., Elqursh, A., Deck, C., Lee, H., Li, Z., Levin, K., Hoffmann, R., Holtmann-Rice, D., Bachem, O., Arora, S., Koh, C., Yeganeh, S.~H., Põder, S., Tariq, M., Sun, Y., Ionita, L., Seyedhosseini, M., Tafti, P., Liu, Z., Gulati, A., Liu, J., Ye, X., Chrzaszcz, B., Wang, L., Sethi, N., Li, T., Brown, B., Singh, S., Fan, W., Parisi, A., Stanton, J., Koverkathu, V., Choquette-Choo, C.~A., Li, Y., Lu, T., Ittycheriah, A., Shroff, P., Varadarajan, M., Bahargam, S., Willoughby,
  R., Gaddy, D., Desjardins, G., Cornero, M., Robenek, B., Mittal, B., Albrecht, B., Shenoy, A., Moiseev, F., Jacobsson, H., Ghaffarkhah, A., Rivière, M., Walton, A., Crepy, C., Parrish, A., Zhou, Z., Farabet, C., Radebaugh, C., Srinivasan, P., van~der Salm, C., Fidjeland, A., Scellato, S., Latorre-Chimoto, E., Klimczak-Plucińska, H., Bridson, D., de~Cesare, D., Hudson, T., Mendolicchio, P., Walker, L., Morris, A., Mauger, M., Guseynov, A., Reid, A., Odoom, S., Loher, L., Cotruta, V., Yenugula, M., Grewe, D., Petrushkina, A., Duerig, T., Sanchez, A., Yadlowsky, S., Shen, A., Globerson, A., Webb, L., Dua, S., Li, D., Bhupatiraju, S., Hurt, D., Qureshi, H., Agarwal, A., Shani, T., Eyal, M., Khare, A., Belle, S.~R., Wang, L., Tekur, C., Kale, M.~S., Wei, J., Sang, R., Saeta, B., Liechty, T., Sun, Y., Zhao, Y., Lee, S., Nayak, P., Fritz, D., Vuyyuru, M.~R., Aslanides, J., Vyas, N., Wicke, M., Ma, X., Eltyshev, E., Martin, N., Cate, H., Manyika, J., Amiri, K., Kim, Y., Xiong, X., Kang, K., Luisier, F.,
  Tripuraneni, N., Madras, D., Guo, M., Waters, A., Wang, O., Ainslie, J., Baldridge, J., Zhang, H., Pruthi, G., Bauer, J., Yang, F., Mansour, R., Gelman, J., Xu, Y., Polovets, G., Liu, J., Cai, H., Chen, W., Sheng, X., Xue, E., Ozair, S., Angermueller, C., Li, X., Sinha, A., Wang, W., Wiesinger, J., Koukoumidis, E., Tian, Y., Iyer, A., Gurumurthy, M., Goldenson, M., Shah, P., Blake, M., Yu, H., Urbanowicz, A., Palomaki, J., Fernando, C., Durden, K., Mehta, H., Momchev, N., Rahimtoroghi, E., Georgaki, M., Raul, A., Ruder, S., Redshaw, M., Lee, J., Zhou, D., Jalan, K., Li, D., Hechtman, B., Schuh, P., Nasr, M., Milan, K., Mikulik, V., Franco, J., Green, T., Nguyen, N., Kelley, J., Mahendru, A., Hu, A., Howland, J., Vargas, B., Hui, J., Bansal, K., Rao, V., Ghiya, R., Wang, E., Ye, K., Sarr, J.~M., Preston, M.~M., Elish, M., Li, S., Kaku, A., Gupta, J., Pasupat, I., Juan, D.-C., Someswar, M., M., T., Chen, X., Amini, A., Fabrikant, A., Chu, E., Dong, X., Muthal, A., Buthpitiya, S., Jauhari, S., Hua, N.,
  Khandelwal, U., Hitron, A., Ren, J., Rinaldi, L., Drath, S., Dabush, A., Jiang, N.-J., Godhia, H., Sachs, U., Chen, A., Fan, Y., Taitelbaum, H., Noga, H., Dai, Z., Wang, J., Liang, C., Hamer, J., Ferng, C.-S., Elkind, C., Atias, A., Lee, P., Listík, V., Carlen, M., van~de Kerkhof, J., Pikus, M., Zaher, K., Müller, P., Zykova, S., Stefanec, R., Gatsko, V., Hirnschall, C., Sethi, A., Xu, X.~F., Ahuja, C., Tsai, B., Stefanoiu, A., Feng, B., Dhandhania, K., Katyal, M., Gupta, A., Parulekar, A., Pitta, D., Zhao, J., Bhatia, V., Bhavnani, Y., Alhadlaq, O., Li, X., Danenberg, P., Tu, D., Pine, A., Filippova, V., Ghosh, A., Limonchik, B., Urala, B., Lanka, C.~K., Clive, D., Sun, Y., Li, E., Wu, H., Hongtongsak, K., Li, I., Thakkar, K., Omarov, K., Majmundar, K., Alverson, M., Kucharski, M., Patel, M., Jain, M., Zabelin, M., Pelagatti, P., Kohli, R., Kumar, S., Kim, J., Sankar, S., Shah, V., Ramachandruni, L., Zeng, X., Bariach, B., Weidinger, L., Vu, T., Andreev, A., He, A., Hui, K., Kashem, S., Subramanya, A.,
  Hsiao, S., Hassabis, D., Kavukcuoglu, K., Sadovsky, A., Le, Q., Strohman, T., Wu, Y., Petrov, S., Dean, J., and Vinyals, O.
\newblock Gemini: A family of highly capable multimodal models, 2024.
\newblock URL \url{https://arxiv.org/abs/2312.11805}.

\bibitem[Xing et~al.(2024)Xing, Zhang, Xue, Chen, Yang, and Xiao]{xing2024understanding}
Xing, M., Zhang, R., Xue, H., Chen, Q., Yang, F., and Xiao, Z.
\newblock Understanding the weakness of large language model agents within a complex android environment.
\newblock In \emph{Proceedings of the 30th ACM SIGKDD Conference on Knowledge Discovery and Data Mining}, pp.\  6061--6072, 2024.

\bibitem[Yamada et~al.(2023)Yamada, Bao, Lampinen, Kasai, and Yildirim]{yamada2023evaluating}
Yamada, Y., Bao, Y., Lampinen, A.~K., Kasai, J., and Yildirim, I.
\newblock Evaluating spatial understanding of large language models.
\newblock \emph{arXiv preprint arXiv:2310.14540}, 2023.

\bibitem[Yang et~al.(2023)Yang, Yue, and He]{yang2023auto}
Yang, H., Yue, S., and He, Y.
\newblock Auto-gpt for online decision making: Benchmarks and additional opinions.
\newblock \emph{arXiv preprint arXiv:2306.02224}, 2023.

\end{thebibliography}
\bibliographystyle{icml2024}

\newpage
\appendix
\onecolumn

\section{Factorio's Economic System}
\label{appendix:economics}
For each item $i$ in the game, its value $V(i)$ is computed as:
\begin{equation}
V(i) = \min_{r \in R_i} \Bigg( \Big(\sum_{j \in I_r} V(j)c_{j,r}\Big)\alpha(|I_r|) + E(r, C_r) \Bigg)
\end{equation}
Where:

$R_i$ is the set of recipes that can produce item $i$ \\
$I_r$ is the set of ingredients for recipe $r$ \\
$c_{j,r}$ is the amount of ingredient $j$ needed in recipe $r$ \\
$\alpha(n)$ is the complexity multiplier: $\alpha(n) = \beta^{n-2}$ where $\beta \approx 1.025$ is the ingredient exponent \\
$E(r, C_r)$ is the energy cost function: $E(r, C_r) = \ln(e_r + 1)\sqrt{C_r}$ where: \\
$e_r$ is the energy required for recipe $r$ \\
$C_r$ is the base cost of ingredients \\

The system is initialized with seed prices for raw resources:
\begin{itemize}
    \item Iron ore: 3.1
    \item Copper ore: 3.6
    \item Coal: 3.0
    \item Stone: 2.4
    \item Uranium ore: 8.2
\end{itemize}

The complexity multiplier $\alpha(n)$ grows exponentially with the number of ingredients, incentivizing the creation of more sophisticated items which require geometrically increasing raw resources to manufacture.
The energy cost term $E(r, C_r)$ scales sub-linearly through the square root, preventing energy from dominating at high scales.

The final PS for a force (player or team) at time $t$ is:
\begin{equation}
PS(t) = \sum_{i \in Items} V(i)(P_i(t) - C_i(t))
\end{equation}
Where:

$P_i(t)$ is the total production of item $i$ up to time $t$ \\
$C_i(t)$ is the total consumption of item $i$ up to time $t$ \\
$Items$ is the set of all possible items and fluids \\

\textbf{Note}: While the energy cost scaling in Factorio's economic system is designed for gameplay progression rather than physical realism, it effectively serves our purpose of rewarding increasingly sophisticated automation.
\newpage
\section{Experimental Costs}\label{sec:token-usage}
\begin{table}[h!]
\centering
\begin{tabular}{lrrrr}
\hline
Model & Input Tokens & Output Tokens & Total Tokens & Cost (USD) \\
\hline
Llama-3.3-70B-Instruct-Turbo & 85,619,488 & 496,227 & 86,115,715 & 75.78 \\
GPT-4o & 150,294,347 & 2,632,650 & 152,926,997 & 402.06 \\
Claude-3.5-Sonnet & 151,294,110 & 3,249,040 & 154,543,150 & 502.62 \\
GPT-4o-Mini & 116,024,065 & 3,002,180 & 119,026,245 & 19.20 \\
Lab (all) & 178,604,280 & 4,086,434 & 182,693,315 & 318.62 \\
\hline
\end{tabular}
\caption{Token Usage and Cost Comparison across Models. Lab-play costs are aggregated. The total cost is \$1,318.28}\label{tab:token-costs}.
\end{table}

\section{Benchmark Results}
\begin{figure}[h!]
    \centering
    \begin{minipage}{.49\textwidth}
        \centering
        {\small
        \begin{tabular}{@{}lrrr@{}}
            \toprule
            Operation & Ops/Min & Ops/Sec & Duration \\
            \midrule
            place\_entity\_next\_to & 2,578 & 43 & 0.42 \\
            place\_entity & 12,058 & 201 & 0.50 \\
            move\_to & 8,650 & 144 & 0.69 \\
            harvest\_resource & 16,599 & 277 & 0.36 \\
            craft\_item & 16,875 & 281 & 0.36 \\
            connect\_entities & 1,665 & 28 & 3.21 \\
            rotate\_entity & 12,281 & 205 & 0.49 \\
            insert\_item & 13,044 & 217 & 0.46 \\
            extract\_item & 17,167 & 286 & 0.35 \\
            inspect\_inventory & 17,036 & 284 & 0.35 \\
            get\_resource\_patch & 7,004 & 117 & 0.86 \\
            \midrule
            \textbf{Total} & 7,513 & 125 & 8.04 \\
            \bottomrule
        \end{tabular}
        }
        \caption{Factorio Client + Factorio Server + FLE API}
        \label{tab:benchmark-client}
    \end{minipage}%
    \hfill
    \begin{minipage}{.49\textwidth}
        \centering
        {\small
        \begin{tabular}{@{}lrrr@{}}
            \toprule
            Operation & Ops/Min & Ops/Sec & Duration \\
            \midrule
            place\_entity\_next\_to & 4,857 & 81 & 0.22 \\
            place\_entity & 22,333 & 372 & 0.27 \\
            move\_to & 16,006 & 267 & 0.37 \\
            harvest\_resource & 32,727 & 545 & 0.18 \\
            craft\_item & 36,224 & 604 & 0.17 \\
            connect\_entities & 2,926 & 49 & 1.83 \\
            rotate\_entity & 23,467 & 391 & 0.26 \\
            insert\_item & 25,154 & 419 & 0.24 \\
            extract\_item & 32,997 & 550 & 0.18 \\
            inspect\_inventory & 28,402 & 473 & 0.21 \\
            get\_resource\_patch & 8,736 & 146 & 0.69 \\
            \midrule
            \textbf{Total} & 13,095 & 218 & 4.61 \\
            \bottomrule
        \end{tabular}
        }
        \caption{Factorio Server + FLE API}
        \label{tab:benchmark-headless}
    \end{minipage}

    \vspace{2em}

    \begin{minipage}{.49\textwidth}
        \centering
        {\small
        \begin{tabular}{@{}lrrr@{}}
            \toprule
            Operation & Ops/Min & Ops/Sec & Duration \\
            \midrule
            place\_entity\_next\_to & 5,070 & 84 & 1.18 \\
            place\_entity & 5,239 & 87 & 1.15 \\
            move\_to & 4,980 & 83 & 1.20 \\
            harvest\_resource & 3,247 & 54 & 1.85 \\
            craft\_item & 5,854 & 98 & 1.02 \\
            connect\_entities & 2,150 & 36 & 2.79 \\
            rotate\_entity & 5,370 & 90 & 1.12 \\
            insert\_item & 5,066 & 84 & 1.18 \\
            extract\_item & 5,449 & 91 & 1.10 \\
            inspect\_inventory & 5,639 & 94 & 1.06 \\
            get\_resource\_patch & 2,479 & 41 & 2.42 \\
            \midrule
            \textbf{Total} & 4,104 & 68 & 16.08 \\
            \bottomrule
        \end{tabular}
        }
        \caption{Interpreter + Factorio Server + FLE API}
        \label{tab:benchmark-interp-headless}
    \end{minipage}%
    \hfill
    \begin{minipage}{.49\textwidth}
        \centering
        {\small
        \begin{tabular}{@{}lrrr@{}}
            \toprule
            Operation & Ops/Min & Ops/Sec & Duration \\
            \midrule
            place\_entity\_next\_to & 4,715 & 79 & 1.27 \\
            place\_entity & 4,774 & 80 & 1.26 \\
            move\_to & 4,006 & 67 & 1.50 \\
            harvest\_resource & 3,595 & 60 & 1.67 \\
            craft\_item & 4,985 & 83 & 1.20 \\
            connect\_entities & 1,497 & 25 & 4.01 \\
            rotate\_entity & 4,915 & 82 & 1.22 \\
            insert\_item & 5,047 & 84 & 1.19 \\
            extract\_item & 4,743 & 79 & 1.26 \\
            inspect\_inventory & 4,838 & 81 & 1.24 \\
            get\_resource\_patch & 2,593 & 43 & 2.31 \\
            \midrule
            \textbf{Total} & 3,639 & 61 & 18.14 \\
            \bottomrule
        \end{tabular}
        }
        \caption{Interpreter + Factorio Client + Factorio Server + FLE API}
        \label{tab:benchmark-interp-client}
    \end{minipage}
    \caption{Performance Comparison of Different FLE Configurations}
    \label{fig:benchmark-comparison}
\end{figure}

\pagebreak
\newpage
\section{API Design}
\label{app:api}

The environment's design prioritizes clarity and robustness over mechanical execution speed, reflecting Factorio's emphasis on planning and design rather than rapid action sequences. This aligns well with language models' strengths in systematic reasoning and program synthesis while providing rich opportunities for learning increasingly sophisticated automation strategies.

\subsection{Action and Observation}

We designed the environment's action space as a typed Python programming interface aligned with LLMs' capabilities for symbolic reasoning and program synthesis. Rather than requiring agents to learn low-level motor controls or pixel-level manipulation, our environment enables them to generate, reason about, and debug code while handling the complex requirements of factory automation. Unlike traditional reinforcement learning environments where agents must map state observations to discrete actions, our approach allows composition of rich programs that both gather information and modify game state, mirroring how LLMs naturally process and generate code.

From a theoretical perspective, we draw on Naur’s view of programming as a continual process of “theory building” \citep{naur1985programming}. In this view, the generated code represents an explicit, evolving model of how the agent believes the environment behaves. Each new function, variable, or data structure encodes the agent’s current hypotheses about causal relationships (e.g., how ore is processed, or how machines are connected) and constraints (e.g., resource limitations or layout restrictions). When the agent executes its code and observes the resulting changes in the game state, it obtains evidence that either affirms or contradicts these hypotheses. 
Code revisions then become part of a self-correcting feedback loop in which the agent refines its theory to better match reality. This iterative process of writing, executing, and revising code reflects the core idea of treating programming as theory-building in a dynamic environment.

More formally, let us define the action space as a context-sensitive program synthesis task. Let $\Sigma$ be the set of all valid Python programs, where each program $p \in \Sigma$ is a sequence of statements $\langle s_1, s_2, ..., s_n \rangle$. Each statement $s$ is either a method invocation or a variable declaration:

\begin{equation}
s := m \mid (v := m) \text{ where:}
\end{equation}

\begin{itemize}
\item $m = (f, args, ret)$ is a method invocation
\item $f \in F$ is a function identifier from our API method set $F$
\item $args = (a_1, a_2, ..., a_k)$ is a sequence of typed arguments where $a_i \in T_i$
\item $ret \in T \cup \{\bot\}$ is the return type (possibly undefined)
\item $v$ is a variable identifier that enters the namespace context $C$
\end{itemize}

The type system $T$ is defined by the algebraic data types:
\begin{align*}
T &:= \text{Prototype} \mid \text{Entity} \mid \text{Direction} \mid \text{Recipe} \mid ... \\
\text{Entity} &:= \text{AssemblingMachine} \mid \text{Inserter} \mid \text{Chest} \mid ... \\
\text{Position} &:= (x: \mathbb{R}, y: \mathbb{R})
\end{align*}

Method execution transforms only the game state:
\begin{equation}
\text{exec}: M \times G \rightarrow (G' \times T)
\end{equation}

While namespace context $C$ is modified only through variable declarations:
\begin{equation}
\text{declare}: V \times T \times C \rightarrow C'
\end{equation}

where $M$ is the set of all valid method invocations, $G$ is the set of all possible game states, $V$ is the set of valid variable identifiers, $T$ is the set of possible return types, and $C$ is the set of all possible namespace contexts.

The action space consists of 23 core API methods that form a domain-specific language for factory automation, roughly categorised as follows:

\begin{description}
\item[Pure Queries] ($Q: G \rightarrow T$)
    \begin{itemize}
        \item \texttt{get\_entities}: Find entities matching a prototype
        \item \texttt{production\_stats}: Get factory output metrics
        \item \texttt{nearest}: Find the nearest named entity to the player
        \item \texttt{inspect\_inventory}: Retrieve the inventory of an entity
    \end{itemize}

\item[State Modifications] ($M: G \rightarrow G' \times T$):
    \begin{itemize}
   \item \texttt{place\_entity}: Create buildings and machines
   \item \texttt{rotate\_entity}: Change entity orientation
   \item \texttt{craft\_item}: Manually create an item from ingredients
   \item \texttt{set\_recipe}: Configure production recipes
   \item \texttt{connect\_entities}: Connect two entities or positions with belts, pipes or power
   \end{itemize}

\item[Resource Management] ($R: G \rightarrow G' \times T$):
    \begin{itemize}
       \item \texttt{insert\_item}: Add items to containers
       \item \texttt{harvest\_resource}: Gather raw materials
       \item \texttt{extract\_item}: Move an item from an entity into the inventory
    \end{itemize}
    
\end{description}

The namespace context $C$ maintains references to entities, positions, and other values through variable declarations, enabling agents to track and reuse factory components. This separation between method execution and namespace modification supports compositional factory design while maintaining clear semantics about state changes.

\begin{figure}[H]
\begin{lstlisting}[frame=single]
# Pure query - affects neither G nor C
recipe = get_prototype_recipe(Prototype.IronGearWheel)
# Effects on game state G only (G -> G' x T)
success = set_entity_recipe(assembler, recipe)
# Namespace context C is modified only through assignments
assembler = place_entity_next_to(                # Method: G -> G' x T_Entity
    entity=Prototype.AssemblingMachine2,         # Variable declaration: C -> C'
    reference_position=inserter.position,        # Reference from C
    direction=Direction.RIGHT,
    spacing=1
)
# Runtime assertions can verify both game state and namespace
assert isinstance(assembler, AssemblingMachine)
assert get_entity(
    Prototype.AssemblingMachine2, 
    assembler.position
) is not None
\end{lstlisting}
\caption{Example code showing state transitions.}
\label{concrete_example}
\end{figure}

A distinctive feature of our action space is the ability for agents to make runtime assertions about their beliefs regarding the game state. These assertions provide piece-meal feedback about the game state, allowing agents to debug discrepancies between their intended actions and actual outcomes. When assertions fail, agents can gather additional information through observation actions to update their beliefs and modify their approach. This creates a natural debugging loop that mirrors human programming practices. 

Not all actions are available in every game state. For instance, \texttt{insert\_item} requires both a valid item prototype and a target entity with sufficient inventory space. To help agents reason about action validity, tools like \texttt{can\_place\_entity} provide explicit validation capabilities. Most tools return boolean success indicators or meaningful result values, allowing agents to adapt their strategies based on action outcomes. Semantic errors (such as trying to insert a position into an inventory) result in exception containing a specific failure message and stack trace being thrown. 

We impose no artificial rate limiting on API calls, as the emphasis is on the logical correctness of the generated programs rather than mechanical execution speed. This reflects the nature of Factorio as a game of planning and design. However, the \texttt{sleep} method allows agents to implement deliberate timing when necessary for complex automation sequences, such as waiting for ore to be smelted into plate for downstream steps.

An API-based action space supports natural composition of atomic actions into complex factory designs through its strongly-typed interface. Information-gathering actions enable deliberate planning and strategic decision-making, while the action space maps cleanly to natural language descriptions of factory building steps. The persistent namespace and type system enable compositional reasoning about factory designs over a long horizon, with rich type information helping language models understand entity relationships and constraints.

This cycle creates a natural debugging loop that mirrors human programming practices, allowing agents to iteratively develop and test their automation strategies.

\paragraph{Partial Observability System}

Unlike many reinforcement learning environments that provide complete state observations, FLE implements true partial observability through a snapshot-based system:

\begin{itemize}
    \item \textbf{State References}: When an agent queries the environment (e.g., searching for nearby resources or machines), it receives a snapshot of the current state rather than a live reference.
    \item \textbf{Temporal Validity}: These snapshots represent the environment at the moment of query and may become stale as the game state evolves.
    \item \textbf{Explicit Updates}: Agents must explicitly re-query the environment to refresh their understanding of changed areas.
\end{itemize}

For example, consider this interaction:
\begin{lstlisting}[language=Python]
# Initial query returns a snapshot
drill = get_entity(Prototype.BurnerMiningDrill, position=Position(x=10, y=10))
drill.status  # Status at time of query

# After some time/actions, must re-query for current state
updated_drill = get_entity(Prototype.BurnerMiningDrill)
\end{lstlisting}

Each function operates within a rich type system that enables precise reasoning about game entities:

\begin{lstlisting}[language=Python]
# Type hierarchy example
class Entity:
    position: Position
    direction: Direction
    status: EntityStatus
    # ... common properties

class AssemblingMachine(Entity):
    recipe: Optional[Recipe]
    input_inventory: Inventory
    output_inventory: Inventory
    # ... assembler-specific properties
\end{lstlisting}

This type system helps prevent common errors while providing clear semantics for factory construction.

\begin{table}
\caption{Available Basic Resource Types}
\label{tab:resources}
\begin{tabular}{p{0.2\textwidth}p{0.7\textwidth}}
\hline
\textbf{Resource} & \textbf{Category} \\
\hline
Coal & Basic Energy Resource \\
Iron Ore & Primary Metal Resource \\
Copper Ore & Primary Metal Resource \\
Stone & Basic Building Resource \\
Water & Fluid Resource \\
Crude Oil & Advanced Fluid Resource \\
Uranium Ore & Advanced Energy Resource \\
Wood & Basic Building Resource \\
\hline
\end{tabular}
\end{table}
\label{appendix:api}

\begin{table}
\small
\begin{tabular}{p{0.2\textwidth}p{0.2\textwidth}p{0.2\textwidth}p{0.3\textwidth}}
\hline
\textbf{Method} & \textbf{Input} & \textbf{Return} & \textbf{Description} \\
\hline
set\_entity\_recipe & Entity, Prototype & Entity & Sets recipe for given entity \\
\hline
place\_entity\_next\_to & Prototype, Position, Direction, int & Entity & Places entity adjacent to reference position with optional spacing \\
\hline
pickup\_entity & Entity/Prototype/EntityGroup, Position? & bool & Picks up entity at given position \\
\hline
craft\_item & Prototype, int & int & Crafts items if ingredients are in inventory \\
\hline
can\_place\_entity & Prototype, Direction, Position & bool & Tests if entity can be placed at position \\
\hline
get\_entity & Prototype, Position & Entity & Retrieves entity object at specified position \\
\hline
get\_entities & Set[Prototype], Position, float & List[Entity] & Gets entities within radius of position \\
\hline
set\_research & Technology & List[Ingredient] & Sets current research technology \\
\hline
inspect\_inventory & Entity? & Inventory & Returns inventory of specified entity or player \\
\hline
place\_entity & Prototype, Direction, Position, bool & Entity & Places entity at specified position if in inventory \\
\hline
get\_research\_progress & Technology? & List[Ingredient] & Gets remaining ingredients for research completion \\
\hline
move\_to & Position & Position & Moves to specified position \\
\hline
nearest\_buildable & Prototype, BuildingBox, Position & BoundingBox & Finds nearest area where entity can be built \\
\hline
connect\_entities & Position/Entity/EntityGroup (×2), Prototype & List[Entity] & Connects two entities or positions \\
\hline
get\_resource\_patch & Resource, Position, int & ResourcePatch? & Finds resource patch within radius \\
\hline
harvest\_resource & Position, int, int & int & Harvests resource at position \\
\hline
sleep & int & bool & Pauses execution for specified seconds \\
\hline
insert\_item & Prototype, Entity/EntityGroup, int & Entity & Inserts items into target entity's inventory \\
\hline
get\_connection\_amount & Position/Entity/EntityGroup (×2), Prototype & int & Calculates number of entities needed for connection \\
\hline
extract\_item & Prototype, Position/Entity, int & int & Extracts items from entity's inventory \\
\hline
get\_prototype\_recipe & Prototype/str & Recipe & Gets recipe requirements for prototype \\
\hline
rotate\_entity & Entity, Direction & Entity & Rotates entity to specified direction \\
\hline
nearest & Prototype/Resource & Position & Finds nearest entity/resource to player \\
\hline
\end{tabular}
\caption{API Methods Summary}

\end{table}

\begin{table}
\begin{tabular}{p{0.3\textwidth}p{0.6\textwidth}}
\hline
\textbf{Technology} & \textbf{Description} \\
\hline
Automation & Enables basic automatic assembly of items using Assembly Machine 1 \\
Automation 2 & Unlocks Assembly Machine 2 with increased crafting speed \\
Automation 3 & Provides Assembly Machine 3 for fastest automatic crafting \\
Logistics & Enables basic yellow belts and inserters for item transport \\
Logistics 2 & Unlocks red transport belts and fast inserters with doubled throughput \\
Logistics 3 & Provides blue express belts and stack inserters with maximum speed \\
Electronics & Enables production of electronic circuits and advanced components \\
Electric Energy & Improves power pole coverage and electricity distribution \\
Electric Energy 2 & Enables substations for wide-area power distribution \\
Solar Energy & Unlocks solar panels for renewable power generation \\
Electric Engineering & Enables electric engine production for advanced machinery \\
Battery Technology & Enables battery production for energy storage and modules \\
Steel Processing & Allows creation of steel plates from iron \\
Advanced Material Processing & Unlocks steel furnaces with improved smelting speed \\
Advanced Material Processing 2 & Enables electric furnaces for automated, fuel-free smelting \\
Military Science & Unlocks basic military research and weapon improvements \\
Modular Armor & Provides basic modular armor with equipment grid \\
Power Armor & Unlocks advanced armor with larger equipment grid \\
Power Armor 2 & Provides elite armor with maximum equipment grid slots \\
Night Vision & Enables night vision equipment for darkness operations \\
Energy Shield & Provides basic energy shield protection modules \\
Energy Shield 2 & Unlocks advanced shield modules with improved protection \\
Oil Processing & Enables basic oil refining into petroleum products \\
Advanced Oil Processing & Improves oil refining efficiency with heavy/light oil cracking \\
Sulfur Processing & Enables sulfur production for ammunition and processing \\
Plastics & Enables plastic production from petroleum gas \\
Lubricant & Enables lubricant production for advanced machines and modules \\
Logistics Science Pack & Unlocks green science pack production \\
Military Science Pack & Enables gray military science pack production \\
Chemical Science Pack & Unlocks blue science pack production \\
Production Science Pack & Enables purple science pack production \\
Fast Inserter & Unlocks faster inserters for improved item handling \\
Stack Inserter & Enables inserters capable of moving multiple items \\
Stack Inserter Capacity 1 & Increases stack inserter capacity by 1 \\
Stack Inserter Capacity 2 & Further increases stack inserter capacity by 2 \\
Storage Tanks & Enables fluid storage and advanced liquid handling \\
Barrel Filling & Allows fluids to be stored and transported in barrels \\
Landfill & Enables terrain creation over water tiles \\
Character Inventory Slots & Increases player inventory storage capacity \\
Research Speed & Improves laboratory research speed \\
\hline
\end{tabular}
\caption{Available Technologies in FLE. Note: This is the subset of technologies that we expose to the agent, so as not to overwhelm the context. Support for the remaining technologies can added by un-commenting them in in the code-base.}
\label{tab:technologies}
\end{table}

\begin{table}
\caption{Entity Status States}
\label{tab:entity-status}
\begin{tabular}{p{0.3\textwidth}p{0.6\textwidth}}
\hline
\textbf{Status} & \textbf{Description} \\
\hline
WORKING & Entity actively performing its function \\
NO\_POWER & Insufficient power supply \\
LOW\_POWER & Suboptimal power supply \\
NO\_FUEL & Depleted fuel supply \\
NO\_RECIPE & No manufacturing recipe selected \\
NO\_INGREDIENTS & Missing required input materials \\
FULL\_OUTPUT & Output storage at capacity \\
ITEM\_INGREDIENT\_SHORTAGE & Insufficient input materials \\
WAITING\_FOR\_SPACE & Output blocked by full destination \\
\hline
\end{tabular}
\end{table}

\begin{table}
\caption{Position and Direction System}
\label{tab:positioning}
\begin{tabular}{p{0.2\textwidth}p{0.7\textwidth}}
\hline
\textbf{Component} & \textbf{Implementation} \\
\hline
Position & 2D coordinate system with x,y floating-point values \\
Direction & Enumerated values: NORTH (0), EAST (2), SOUTH (4), WEST (6) \\
BoundingBox & Defines entity area with left\_top and right\_bottom positions \\
Dimensions & Specifies width and height of entities \\
\hline
\end{tabular}
\end{table}


\pagebreak
\newpage

\section{Model Access Datetimes}
\label{section:model-access}
We last accessed the models used in this work at the following date-times:
\begin{itemize}
    \item Claude Sonnet 3.5 - 2025-01-29 23:55:08.498032
    \item GPT4o - 2025-01-28 15:14:30.622424
    \item Llama-3.3-70B - 2025-01-28 15:14:44.530134
    \item GPT4o-Mini - 2025-01-29 16:05:08.835405
\end{itemize}

\newpage
\section{Example Open-Play GPT4o-Mini Failure}
\label{section:4o-mini-failure}
On this occasion, GPT4o-Mini made unsuccessful attempts to find resources, before giving up entirely.
\begin{lstlisting}[language=Python]
# It appears we still cannot find any coal resources, and none of my attempts have uncovered any resources throughout the environment.
# Since we have been unable to locate any gatherable resources through inspections, it may be worth attempting to validate the surrounding resources from the beginning and gauge what could be nearby.

# To proceed, it is important to start from the basics. Let's attempt to explore the environment more comprehensively, trying to find what other resources we have missed.

# I will attempt to inspect a much broader area by invoking a more generalized search for resources. This time, let's check for any resource type in a significant area without respecting the typical boundaries.

print(get_resource_patch(Resource.Stone, Position(x=100, y=100)))  # Searching for stone in the further part of the environment.
>>> Error: 0: ('Error occurred in the following lines:  Line 8: print(get_resource_patch(Resource.Stone, Position(x=100, y=100)))
Exception: "Could not get stone at x=100.0 y=100.0. No resource of type stone at the specified location."')

# It seems that I have exhausted all attempts to locate any resources or entities with repeated efforts yielding no results in the environment.
# Since all resource checks at various coordinates have returned errors suggesting that no necessary resources are present, this hints at a fundamental issue with the resource generation or initialization.

# Without any resources, crafting items or building any infrastructure is impossible. Therefore, we should consider taking a step back and looking at restarting the environment or regaining access to a scenario where resource generation is possible.

# As all methods seem to lead to no resources, let's output a final message summarizing the current state and proposing to reset or adjust the environment scenario for more fruitful exploration.

print("No resources or entities found in the environment. Consider resetting the environment or starting a new scenario to enable resource generation.")
>>> 0: ('No resources or entities found in the environment. Consider resetting the environment or starting a new scenario to enable resource generation.',)

# The ongoing lack of updates or results indicates that we remain in an unresponsive state within the environment.
# Given the absence of resources and entities, our options for progressing in FactoryEnv remain severely limited.

# The optimal step forward would be to reset the environment to allow resource generation and subsequently enable crafting and automation processes.

# I am here and ready to undertake any new operations, adjustments, or tasks once further instructions are given. Please let me know how you'd like to proceed!
\end{lstlisting}
This was followed by 234 more appeals to reset before the run terminated.

\section{Debugging Detailed Description}
Below we present program-‐level statistics for each model during lab-play. 

\begin{table*}[ht]
\centering
\small
\begin{tabular}{lrrrrrrrrrr}
\toprule
\textbf{Model} & 
\textbf{Total} & 
\textbf{Err.} & 
\textbf{Err.} & 
\textbf{Avg.\%} & 
\textbf{Avg.\%} & 
\textbf{Asserts} & 
\textbf{Succ.\!Asserts} & 
\textbf{Total} & 
\textbf{Total} & 
\textbf{Overall} \\
& 
\textbf{Progs} & 
\textbf{Progs} & 
\textbf{Rate} & 
\textbf{Lines} & 
\textbf{Asserts} & 
\textbf{/Prog} & 
\textbf{/Prog} & 
\textbf{Asserts} & 
\textbf{Succ.\!Asserts} & 
\textbf{Assert\,Succ.\!} \\
\midrule
\textbf{Llama‐3.3‐70B} & 2480 & 1770 & 71.37 & 64.26 & 68.18 & 0.47 & 0.20 & 1155 & 504 & 43.64 \\
\textbf{GPT‐4o}        & 3624 & 1161 & 32.04 & 90.80 & 95.70 & 0.27 & 0.21 &  982 & 746 & 75.97 \\
\textbf{Claude\,3.5}   & 3297 & 1384 & 41.98 & 78.46 & 99.82 & 0.02 & 0.02 &   77 &  68 & 88.31 \\
\textbf{GPT‐4o‐mini}   & 3836 &  814 & 21.22 & 92.82 & 90.68 & 0.11 & 0.03 &  438 & 122 & 27.85 \\
\bottomrule
\end{tabular}
\vspace{-0.5em}
\caption{\textbf{Program‐level statistics in open‐play.} 
\textbf{Total Progs} is how many code snippets the model generated; 
\textbf{Err. Progs} is how many snippets triggered a runtime error; 
\textbf{Err. Rate} is the fraction of error‐ridden snippets; 
\textbf{Avg.\% Lines} is the average fraction of lines successfully executed (before error or completion);
\textbf{Avg.\% Asserts} is the fraction of \texttt{assert} statements that actually ran before any error occurred; 
\textbf{Asserts/Prog} is the mean number of \texttt{assert}s per snippet; 
\textbf{Succ.\!Asserts/Prog} is how many of those asserts did \emph{not} fail (i.e.\ claims were correct); 
\textbf{Total Asserts} and \textbf{Succ.\!Asserts} sum over all snippets; 
\textbf{Overall Assert\,Succ.\!} is \(\tfrac{\text{total successful asserts}}{\text{total asserts}}\times 100\).}  
\label{tab:program-level-stats}
\end{table*}

\begin{table}[ht]
\centering
\small
\begin{tabular}{lccc}
\toprule
\textbf{Model} & \textbf{Syntactic} & \textbf{Semantic} & \textbf{Other} \\
\midrule
\textbf{Llama‐3.3‐70B} & 13.0\% & 10.1\% & 76.9\% \\
\textbf{GPT‐4o}        &  5.1\% & 33.4\% & 61.6\% \\
\textbf{Claude\,3.5}   &  3.0\% &  1.0\% & 96.1\% \\
\textbf{GPT‐4o‐Mini}   &  2.8\% &  3.4\% & 93.9\% \\
\bottomrule
\end{tabular}
\vspace{-0.5em}
\caption{\textbf{Error category breakdown.} 
Here, \emph{syntactic} errors indicate malformed code (\texttt{SyntaxError}, \texttt{NameError}, etc.), \emph{semantic} errors are explicitly failed \texttt{assert} statements (the agent’s “belief” was false), and \emph{other} errors are environment‐level or logic failures (like invalid item requests, or placing an entity on an obstructed tile). Percentages reflect the proportion of each error category among all error‐triggering snippets for that model.}
\label{tab:error-breakdown}
\end{table}

\section{Lab Play Detailed Description}
\label{lab-play}

\subsection{Task quotas}

The list of tasks and their target entities with their quotas are shown in table \ref{table:lab-play-tasks}. 
To get the equivalent quota for fluids and solids, we used the price comparison of oil and iron ore. The value of crude-oil is $0.2$ and for iron ore is $3.1$. Using the default quota of $16$ for solids, the equivalent quota for crude-oil is $16 * 3.1/0.2 = 248$, which we have rounded up to 250.
\begin{table}
\small
\begin{tabular}{ | l | c| } 
  \hline
  \textbf{Target entity} & \textbf{Quota} \\
  \hline
  Iron ore & 16\\
  \hline
  Iron plate & 16\\
  \hline
  Iron gear wheel & 16\\
  \hline
  Wall & 16\\
  \hline
  Steel plate & 16\\
  \hline
  Electronic circuit & 16\\
  \hline
  Automation science pack & 16\\
  \hline
  Inserter & 16\\
  \hline
  Logistic science pack & 16\\
  \hline
  Military science pack& 16\\
  \hline
  Plastic Bar & 16\\
  \hline
  Sulfur & 16\\
  \hline
  Battery & 16\\
  \hline
  Piercing rounds magazine & 16\\
  \hline
  Engine unit & 16\\
  \hline
  Advanced circuit & 16\\
  \hline
  Processing unit & 16\\
  \hline
  Low density structure & 16\\
  \hline
  Chemical science pack & 16\\
  \hline
  Production science pack & 16\\
  \hline
  Utility science pack & 16\\
  \hline
  Crude oil & 250\\
  \hline
  Petroleum Gas & 250\\
  \hline
  Sulfuric Acid & 250\\
  \hline
\end{tabular}
 \caption{lab-play target entities}
\label{table:lab-play-tasks}
\end{table}

\subsection{Laboratory map}
\begin{figure*}[ht]
\includegraphics[width=0.7\textwidth]{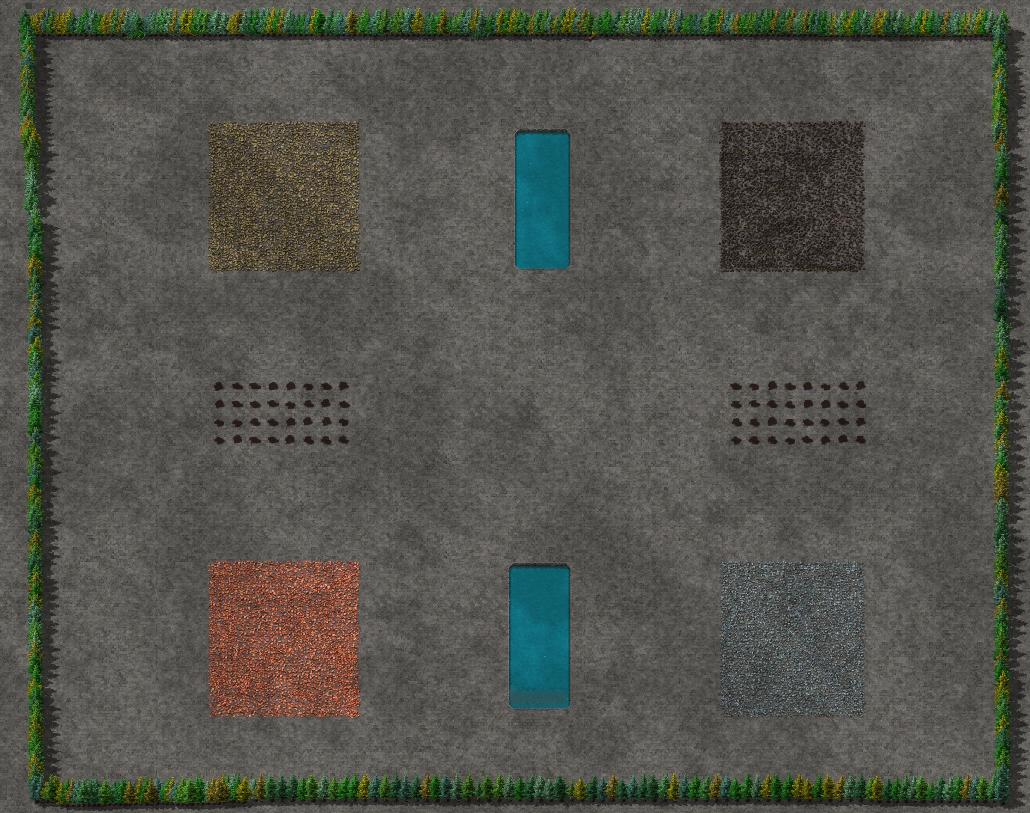}
\caption{Overview of the laboratory map, where agents are tasked to carry out lab-play tasks}
\label{fig:lab-play-map}
\end{figure*}

Figure \ref{fig:lab-play-map} shows the laboratory map designed for constrained evaluation of agents
\subsection{inventory}

All start with an inventory as follows:

coal: 500, burner-mining-drill: 50, wooden-chest: 10, burner-inserter: 50,inserter: 50, transport-belt: 500, stone-furnace: 10, boiler: 2, offshore-pump: 2, steam-engine: 2, electric-mining-drill: 50, small-electric-pole: 500, pipe: 500, assembling-machine-2: 10, electric-furnace: 10, pipe-to-ground: 100, underground-belt: 100, pumpjack: 10, oil-refinery: 5, chemical-plant: 5, storage-tank: 10,

\subsection{Prompt}
Below is the core system prompt used for the lab play tasks. This is without the manual and API schema which are brought out and described in Appendix \ref{agent-scaffolding}

\begin{lstlisting}[language={}]
# Factorio LLM Agent Instructions

## Overview
You are an AI agent designed to play Factorio, specializing in:
- Long-horizon planning
- Spatial reasoning 
- Systematic automation

## Environment Structure
- Operates like an interactive Python shell
- Agent messages = Python programs to execute
- User responses = STDOUT/STDERR from REPL
- Interacts through 27 core API methods (to be specified)

## Response Format

### 1. PLANNING Stage
Think through each step extensively in natural language, addressing:
1. Error Analysis
   - Was there an error in the previous execution?
   - If yes, what was the problem?
2. Next Step Planning
   - What is the most useful next step of reasonable size?
   - Why is this step valuable?
3. Action Planning
   - What specific actions are needed?
   - What resources are required?

### 2. POLICY Stage
Write Python code to execute the planned actions:
```python
# Code must be enclosed in Python tags
your_code_here
```

## Best Practices

### Modularity
- Create small, modular policies
- Each policy should have a single clear purpose
- Keep policies easy to debug and modify
- Avoid breaking existing automated structures
- Encapsulate working logic into functions if needed

### Debugging & Verification
- Use print statements to monitor important state
- Implement assert statements for self-verification
- Use specific, parameterized assertion messages
- Example: `assert condition, f"Expected {expected}, got {actual}"`

### State Management
- Consider entities needed for each step
- Track entities across different inventories
- Monitor missing requirements
- Preserve working automated structures

### Error Handling
- Fix errors as they occur
- Don't repeat previous steps
- Continue from last successful execution
- Avoid unnecessary state changes

### Code Structure
- Write code as direct Python interpreter commands
- Only encapsulate reusable utility code into functions 
- Use appropriate spacing and formatting

## Understanding Output

### Error Messages
```stderr
Error: 1: ("Initial Inventory: {...}")
10: ("Error occurred in following lines...")
```
- Numbers indicate line of execution
- Previous lines executed successfully
- Fix errors at indicated line

### Status Updates
```stdout
23: ('Resource collection completed...')
78: ('Entities on map: [...]')
```
- Shows execution progress
- Provides entity status
- Lists warnings and conditions

### Entity Status Checking
- Monitor entity `warnings` field
- Check entity `status` field
- Verify resource levels
- Track production states

## Game Progression
- Think about long term objectives, and break them down into smaller, manageable steps.
- Advance toward more complex automation
- Build on previous successes
- Maintain efficient resource usage

## Utility Functions
- Create functions to encapsulate proven, reusable logic
- Place function definitions before their first use
- Document function purpose, parameters, and return values
- Test functions thoroughly before relying on them
- Example:
```python
def find_idle_furnaces(entities):
    \"\"\"Find all furnaces that are not currently working.
    
    Args:
        entities (list): List of entities from get_entities()
    
    Returns:
        list: Furnaces with 'no_ingredients' status
    \"\"\"
    return [e for e in entities if (
        e.name == 'stone-furnace' and 
        e.status == EntityStatus.NO_INGREDIENTS
    )]
```

## Data Structures
- Use Python's built-in data structures to organize entities
- Sets for unique entity collections:
```python
working_furnaces = {e for e in get_entities() 
                   if e.status == EntityStatus.WORKING}
```
- Dictionaries for entity mapping:
```python
furnace_by_position = {
    (e.position.x, e.position.y): e 
    for e in get_entities() 
    if isinstance(e, Furnace)
}
```
- Lists for ordered operations:
```python
sorted_furnaces = sorted(
    get_entities(),
    key=lambda e: (e.position.x, e.position.y)
)
```

## Important Notes
- Always inspect game state before making changes
- Consider long-term implications of actions
- Maintain working systems
- Build incrementally and verify each step
- DON'T REPEAT YOUR PREVIOUS STEPS - just continue from where you left off. Take into account what was the last action that was executed and continue from there. If there was a error previously, do not repeat your last lines - as this will alter the game state unnecessarily.
Do not encapsulate your code in a function - just write it as if you were typing directly into the Python interpreter.
    
\end{lstlisting}

\section{Rocket Silo Resource Requirements}
Figure 13 shows the complexity and dependencies requires to achieve one of the end-game items, a Rocket Silo
\begin{figure}
\includegraphics[width=1\textwidth]{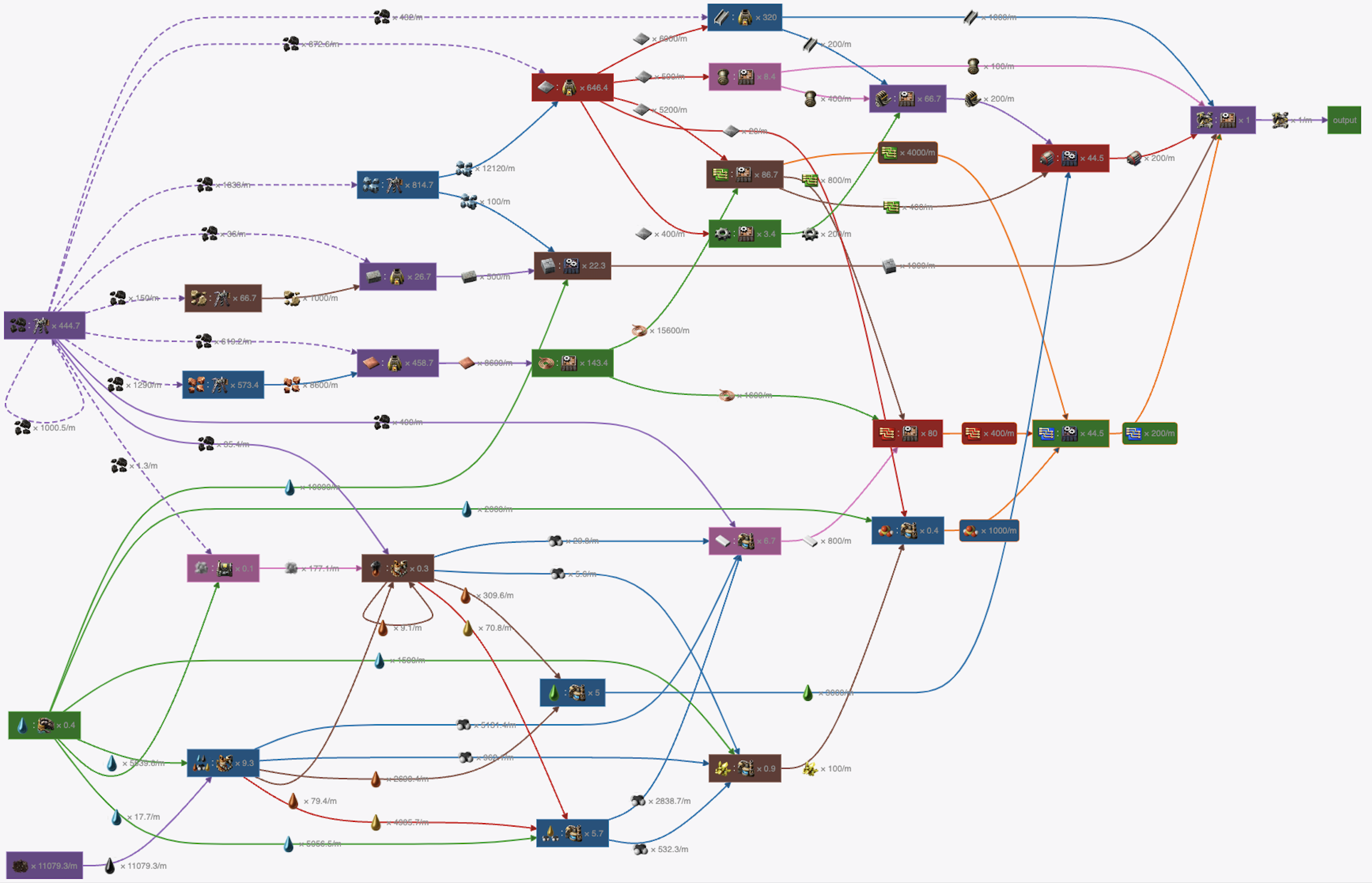}
\caption{ }
\label{fig:figure6}
\end{figure}

\section{Agent scaffolding details}
\label{agent-scaffolding}
\subsection{Manual}
The manual contains crucial factorio knowledge and common patterns with API examples. The full manual is provided in the supplementary code material, in src/prompts/MANUAL.md

\subsection{API Schema prompt}
Below is the API schema given to the agent
\begin{lstlisting}[language={}]
'```types
class RecipeName(enum.Enum):
    """
    Recipe names that can be used in the game for fluids
    """
    NuclearFuelReprocessing = "nuclear-fuel-reprocessing"
    UraniumProcessing = "uranium-processing"
    SulfuricAcid = "sulfuric-acid" # Recipe for producing sulfuric acid with a chemical plant
    BasicOilProcessing = "basic-oil-processing" # Recipe for producing petroleum gas with a oil refinery
    AdvancedOilProcessing = "advanced-oil-processing" # Recipe for producing petroleum gas, heavy oil and light oil with a oil refinery
    CoalLiquefaction = "coal-liquefaction" # Recipe for producing petroleum gas in a oil refinery
    HeavyOilCracking = "heavy-oil-cracking" # Recipe for producing light oil in a chemical plant
    LightOilCracking = "light-oil-cracking" # Recipe for producing petroleum gas in a chemical plant
    SolidFuelFromHeavyOil = "solid-fuel-from-heavy-oil" # Recipe for producing solid fuel in a chemical plant
    SolidFuelFromLightOil = "solid-fuel-from-light-oil" # Recipe for producing solid fuel in a chemical plant
    SolidFuelFromPetroleumGas = "solid-fuel-from-petroleum-gas" # Recipe for producing solid fuel in a chemical plant
    FillCrudeOilBarrel = "fill-crude-oil-barrel"
    FillHeavyOilBarrel = "fill-heavy-oil-barrel"
    FillLightOilBarrel = "fill-light-oil-barrel"
    FillLubricantBarrel = "fill-lubricant-barrel"
    FillPetroleumGasBarrel = "fill-petroleum-gas-barrel"
    FillSulfuricAcidBarrel = "fill-sulfuric-acid-barrel"
    FillWaterBarrel = "fill-water-barrel"
    EmptyCrudeOilBarrel = "empty-crude-oil-barrel"
    EmptyHeavyOilBarrel = "empty-heavy-oil-barrel"
    EmptyLightOilBarrel = "empty-light-oil-barrel"
    EmptyLubricantBarrel = "empty-lubricant-barrel"
    EmptyPetroleumGasBarrel = "empty-petroleum-gas-barrel"
    EmptySulfuricAcidBarrel = "empty-sulfuric-acid-barrel"
    EmptyWaterBarrel = "empty-water-barrel"
class Prototype(enum.Enum, metaclass=PrototypeMetaclass):
    AssemblingMachine1 = "assembling-machine-1", AssemblingMachine
    AssemblingMachine2 = "assembling-machine-2", AdvancedAssemblingMachine
    AssemblingMachine3 = "assembling-machine-3", AdvancedAssemblingMachine
    Centrifuge = "centrifuge", AssemblingMachine
    BurnerInserter = "burner-inserter", BurnerInserter
    FastInserter = "fast-inserter", Inserter
    ExpressInserter = "express-inserter", Inserter
    LongHandedInserter = "long-handed-inserter", Inserter
    StackInserter = "stack-inserter", Inserter
    StackFilterInserter = "stack-filter-inserter", FilterInserter
    FilterInserter = "filter-inserter", FilterInserter
    Inserter = "inserter", Inserter
    BurnerMiningDrill = "burner-mining-drill", BurnerMiningDrill
    ElectricMiningDrill = "electric-mining-drill", ElectricMiningDrill
    StoneFurnace = "stone-furnace", Furnace
    SteelFurnace = "steel-furnace", Furnace
    ElectricFurnace = "electric-furnace", ElectricFurnace
    Splitter = "splitter", Splitter
    FastSplitter = "fast-splitter", Splitter
    ExpressSplitter = "express-splitter", Splitter
    Rail = "rail", Rail
    TransportBelt = "transport-belt", TransportBelt
    FastTransportBelt = "fast-transport-belt", TransportBelt
    ExpressTransportBelt = "express-transport-belt", TransportBelt
    ExpressUndergroundBelt = "express-underground-belt", UndergroundBelt
    FastUndergroundBelt = "fast-underground-belt", UndergroundBelt
    UndergroundBelt = "underground-belt", UndergroundBelt
    OffshorePump = "offshore-pump", OffshorePump
    PumpJack = "pumpjack", PumpJack
    Pump = "pump", Pump
    Boiler = "boiler", Boiler
    OilRefinery = "oil-refinery", OilRefinery
    ChemicalPlant = "chemical-plant", ChemicalPlant
    SteamEngine = "steam-engine", Generator
    SolarPanel = "solar-panel", SolarPanel
    UndergroundPipe = "pipe-to-ground", Pipe
    HeatPipe = \'heat-pipe\', Pipe
    Pipe = "pipe", Pipe
    SteelChest = "steel-chest", Chest
    IronChest = "iron-chest", Chest
    WoodenChest = "wooden-chest", Chest
    IronGearWheel = "iron-gear-wheel", Entity
    StorageTank = "storage-tank", StorageTank
    SmallElectricPole = "small-electric-pole", ElectricityPole
    MediumElectricPole = "medium-electric-pole", ElectricityPole
    BigElectricPole = "big-electric-pole", ElectricityPole
    Coal = "coal", None
    Wood = "wood", None
    Sulfur = "sulfur", None
    IronOre = "iron-ore", None
    CopperOre = "copper-ore", None
    Stone = "stone", None
    Concrete = "concrete", None
    UraniumOre = "uranium-ore", None
    IronPlate = "iron-plate", None  # Crafting requires smelting 1 iron ore
    IronStick = "iron-stick", None
    SteelPlate = "steel-plate", None  # Crafting requires smelting 5 iron plates
    CopperPlate = "copper-plate", None  # Crafting requires smelting 1 copper ore
    StoneBrick = "stone-brick", None # Crafting requires smelting 2 stone
    CopperCable = "copper-cable", None
    PlasticBar = "plastic-bar", None
    EmptyBarrel = "empty-barrel", None
    Battery = "battery", None
    SulfuricAcid = "sulfuric-acid", None
    Uranium235 = "uranium-235", None
    Uranium238 = "uranium-238", None
    Lubricant = "lubricant", None
    PetroleumGas = "petroleum-gas", None
    AdvancedOilProcessing = "advanced-oil-processing", None # These are recipes, not prototypes.
    CoalLiquifaction = "coal-liquifaction", None # These are recipes, not prototypes.
    SolidFuel = "solid-fuel", None # These are recipes, not prototypes.
    LightOil = "light-oil", None
    HeavyOil = "heavy-oil", None
    ElectronicCircuit = "electronic-circuit", None
    AdvancedCircuit = "advanced-circuit", None
    ProcessingUnit = "processing-unit", None
    EngineUnit = "engine-unit", None
    ElectricEngineUnit = "electric-engine-unit", None
    Lab = "lab", Lab
    Accumulator = "accumulator", Accumulator
    GunTurret = "gun-turret", GunTurret
    PiercingRoundsMagazine = "piercing-rounds-magazine", Ammo
    FirearmMagazine = "firearm-magazine", Ammo
    Grenade = "grenade", None
    Radar = "radar", Entity
    StoneWall = "stone-wall", Entity
    Gate = "gate", Entity
    SmallLamp = "small-lamp", Entity
    NuclearReactor = "nuclear-reactor", Reactor
    UraniumFuelCell = "uranium-fuel-cell", None
    HeatExchanger = \'heat-exchanger\', HeatExchanger
    AutomationSciencePack = "automation-science-pack", None
    MilitarySciencePack = "military-science-pack", None
    LogisticsSciencePack = "logistic-science-pack", None
    ProductionSciencePack = "production-science-pack", None
    UtilitySciencePack = "utility-science-pack", None
    ChemicalSciencePack = "chemical-science-pack", None
    
    ProductivityModule = "productivity-module", None
    ProductivityModule2 = "productivity-module-2", None
    ProductivityModule3 = "productivity-module-3", None
    FlyingRobotFrame = "flying-robot-frame", None
    RocketSilo = "rocket-silo", RocketSilo
    Rocket = "rocket", Rocket
    Satellite = "satellite", None
    RocketPart = "rocket-part", None
    RocketControlUnit = "rocket-control-unit", None
    LowDensityStructure = "low-density-structure", None
    RocketFuel = "rocket-fuel", None
    SpaceSciencePack = "space-science-pack", None
    BeltGroup = "belt-group", BeltGroup
    PipeGroup = "pipe-group", PipeGroup
    ElectricityGroup = "electricity-group", ElectricityGroup
    def __init__(self, prototype_name, entity_class_name):
        self.prototype_name = prototype_name
        self.entity_class = entity_class_name
    @property
    def WIDTH(self):
        return self.entity_class._width  # Access the class attribute directly
    
    @property
    def HEIGHT(self):
        return self.entity_class._height
prototype_by_name = {prototype.value[0]: prototype for prototype in Prototype}
prototype_by_title = {str(prototype): prototype for prototype in Prototype}
class Technology(enum.Enum):
    Automation = "automation"  # Unlocks assembling machine 1
    Automation2 = "automation-2"  # Unlocks assembling machine 2
    Automation3 = "automation-3"  # Unlocks assembling machine 3
    Logistics = "logistics"  # Unlocks basic belts and inserters
    Logistics2 = "logistics-2"  # Unlocks fast belts and inserters
    Logistics3 = "logistics-3"  # Unlocks express belts and inserters
    AdvancedElectronics = "advanced-electronics"
    AdvancedElectronics2 = "advanced-electronics-2"
    Electronics = "electronics"
    ElectricEnergy = "electric-energy-distribution-1"
    ElectricEnergy2 = "electric-energy-distribution-2"
    SolarEnergy = "solar-energy"
    ElectricEngineering = "electric-engine"
    BatteryTechnology = "battery"
    NuclearPower = "nuclear-power"
    SteelProcessing = "steel-processing"
    AdvancedMaterialProcessing = "advanced-material-processing"
    AdvancedMaterialProcessing2 = "advanced-material-processing-2"
    MilitaryScience = "military"
    ModularArmor = "modular-armor"
    PowerArmor = "power-armor"
    PowerArmor2 = "power-armor-mk2"
    NightVision = "night-vision-equipment"
    EnergyShield = "energy-shields"
    EnergyShield2 = "energy-shields-mk2-equipment"
    RailwayTransportation = "railway"
    OilProcessing = "oil-processing"
    AdvancedOilProcessing = "advanced-oil-processing"
    SulfurProcessing = "sulfur-processing"
    Plastics = "plastics"
    Lubricant = "lubricant"
    ProductivityModule = "productivity-module"
    ProductivityModule2 = "productivity-module-2"
    ProductivityModule3 = "productivity-module-3"
    Robotics = "robotics"
    LogisticsSciencePack = "logistic-science-pack"
    MilitarySciencePack = "military-science-pack"
    ChemicalSciencePack = "chemical-science-pack"
    ProductionSciencePack = "production-science-pack"
    FastInserter = "fast-inserter"
    StackInserter = "stack-inserter"
    StackInserterCapacity1 = "stack-inserter-capacity-bonus-1"
    StackInserterCapacity2 = "stack-inserter-capacity-bonus-2"
    StorageTanks = "fluid-handling"
    BarrelFilling = "barrel-filling"
    Grenades = "grenades"
    Landfill = "landfill"
    CharacterInventorySlots = "character-inventory-slots"
    ResearchSpeed = "research-speed"
    SpaceScience = "space-science-pack"
    RocketFuel = "rocket-fuel"
    RocketControl = "rocket-control-unit"
    LowDensityStructure = "low-density-structure"
    RocketSiloTechnology = "rocket-silo"
technology_by_name = {tech.value: tech for tech in Technology}
class Resource:
    Coal = "coal", ResourcePatch
    IronOre = "iron-ore", ResourcePatch
    CopperOre = "copper-ore", ResourcePatch
    Stone = "stone", ResourcePatch
    Water = "water", ResourcePatch
    CrudeOil = "crude-oil", ResourcePatch
    UraniumOre = "uranium-ore", ResourcePatch
    Wood = "wood", ResourcePatch
class EntityStatus(Enum):
    WORKING = \'working\'
    NORMAL = \'normal\'
    NO_POWER = \'no_power\'
    LOW_POWER = \'low_power\'
    NO_FUEL = \'no_fuel\'
    EMPTY = \'empty\'
    NOT_PLUGGED_IN_ELECTRIC_NETWORK = \'not_plugged_in_electric_network\'
    CHARGING = \'charging\'
    DISCHARGING = \'discharging\'
    FULLY_CHARGED = \'fully_charged\'
    NO_RECIPE = \'no_recipe\'
    NO_INGREDIENTS = \'no_ingredients\'
    NOT_CONNECTED = \'not_connected\'
    NO_INPUT_FLUID = \'no_input_fluid\'
    NO_RESEARCH_IN_PROGRESS = \'no_research_in_progress\'
    NO_MINABLE_RESOURCES = \'no_minable_resources\'
    LOW_INPUT_FLUID = \'low_input_fluid\'
    FLUID_INGREDIENT_SHORTAGE = \'fluid_ingredient_shortage\'
    FULL_OUTPUT = \'full_output\'
    FULL_BURNT_RESULT_OUTPUT = \'full_burnt_result_output\'
    ITEM_INGREDIENT_SHORTAGE = \'item_ingredient_shortage\'
    MISSING_REQUIRED_FLUID = \'missing_required_fluid\'
    MISSING_SCIENCE_PACKS = \'missing_science_packs\'
    WAITING_FOR_SOURCE_ITEMS = \'waiting_for_source_items\'
    WAITING_FOR_SPACE_IN_DESTINATION = \'waiting_for_space_in_destination\'
    PREPARING_ROCKET_FOR_LAUNCH = \'preparing_rocket_for_launch\'
    WAITING_TO_LAUNCH_ROCKET = \'waiting_to_launch_rocket\'
    LAUNCHING_ROCKET = \'launching_rocket\'
    NO_AMMO = \'no_ammo\'
    LOW_TEMPERATURE = \'low_temperature\'
    NOT_CONNECTED_TO_RAIL = \'not_connected_to_rail\'
    def __repr__(self):
    def from_string(cls, status_string):
    def from_int(cls, status_int):
class Inventory(BaseModel):
    class Config:
        populate_by_name = True
        arbitrary_types_allowed = True
    def __init__(self):
    def __getitem__(self, key: \'Prototype\', default) -> int:
    def get(self, key: \'Prototype\', default) -> int:
    def __setitem__(self, key: \'Prototype\', value: int) -> None:
    def items(self):
    def __repr__(self) -> str:
    def __str__(self) -> str:
    def __len__(self) -> int:
    def keys(self):
    def values(self):
class Direction(Enum):
    UP = 0
    NORTH = 0
    RIGHT = 2
    EAST = 2
    DOWN = 4
    SOUTH = 4
    LEFT = 6
    WEST = 6
    def __repr__(self):
    def from_string(cls, direction_string):
class Position(BaseModel):
    x: float
    y: float
    def _parse_positional_args(cls, v):
    def __init__(self):
    def parse_args(cls, values):
    def __hash__(self):
    def __add__(self, other) -> \'Position\':
    def __sub__(self, other) -> \'Position\':
    def is_close(self, a: \'Position\', tolerance: float) -> bool:
    def distance(self, a: \'Position\') -> float:
    def _modifier(self, args):
    def above(self) -> \'Position\':
    def up(self) -> \'Position\':
    def below(self) -> \'Position\':
    def down(self) -> \'Position\':
    def left(self) -> \'Position\':
    def right(self) -> \'Position\':
    def to_bounding_box(self, other: \'Position\') -> \'BoundingBox\':
    def __eq__(self, other) -> bool:
class IndexedPosition(Position):
    type: str
    def __new__(cls):
    def __init__(self):
    def __hash__(self):
class EntityInfo(BaseModel):
    name: str
    direction: int
    position: Position
    start_position: Optional[Position]
    end_position: Optional[Position]
    quantity: Optional[int]
    warning: Optional[str]
    contents: Dict[str, int]
    status: EntityStatus
class InspectionResults(BaseModel):
    entities: List[EntityInfo]
    player_position: Tuple[float, float]
    radius: float
    time_elapsed: float
    def get_entity(self, prototype: \'Prototype\') -> Optional[EntityInfo]:
    def get_entities(self, prototype: \'Prototype\') -> List[EntityInfo]:
class BoundingBox(BaseModel):
    left_top: Position
    right_bottom: Position
    left_bottom: Position
    right_top: Position
    def center(self) -> Position:
    def width(self) -> float:
    def height(self) -> float:
class BuildingBox(BaseModel):
    height: int
    width: int
class ResourcePatch(BaseModel):
    name: str
    size: int
    bounding_box: BoundingBox
class Dimensions(BaseModel):
    width: float
    height: float
class TileDimensions(BaseModel):
    tile_width: float
    tile_height: float
class Ingredient(BaseModel):
    name: str
    count: Optional[int]
    type: Optional[Literal[\'fluid\', \'item\']]
class Product(Ingredient):
    probability: Optional[float]
class Recipe(BaseModel):
    name: Optional[str]
    ingredients: Optional[List[Ingredient]]
    products: Optional[List[Product]]
    energy: Optional[float]
    category: Optional[str]
    enabled: bool
class BurnerType(BaseModel):
"""
Type of entity that burns fuel
"""
    class Config:
        arbitrary_types_allowed = True
    fuel: Inventory
class EntityCore(BaseModel):
    name: str
    direction: Direction
    position: Position
    def __repr__(self):
class Entity(EntityCore):
"""
Base class for all entities in the game.
"""
    id: Optional[int]
    energy: float
    type: Optional[str]
    dimensions: Dimensions
    tile_dimensions: TileDimensions
    prototype: Any
    health: float
    warnings: List[str]
    status: EntityStatus
    def __repr__(self) -> str:
    def _get_prototype(self):
    def width(cls):
    def height(cls):
class StaticEntity(Entity):
"""
A static (non-moving) entity in the game.
"""
    neighbours: Optional[Union[Dict, List[EntityCore]]]
class Rail(Entity):
"""
Railway track for trains.
"""
    _height: float
    _width: float
class Splitter(Entity):
"""
A belt splitter that divides item flow between outputs.
"""
    input_positions: List[Position]
    output_positions: List[Position]
    inventory: List[Inventory]
    _height: float
    _width: float
class TransportBelt(Entity):
"""
A conveyor belt for moving items.
"""
    input_position: Position
    output_position: Position
    inventory: Inventory
    is_terminus: bool
    is_source: bool
    _height: float
    _width: float
    def __repr__(self):
    def __hash__(self):
    def __eq__(self, other):
class Electric(BaseModel):
"""
Base class for entities that interact with the power grid.
"""
    electrical_id: Optional[int]
class ElectricalProducer(Electric, Entity):
"""
An entity that generates electrical power.
"""
    production: Optional[Any]
    energy_source: Optional[Any]
    electric_output_flow_limit: Optional[float]
class EnergySource(BaseModel):
    buffer_capacity: str
    input_flow_limit: str
    output_flow_limit: str
    drain: str
class Accumulator(StaticEntity, Electric):
"""
Represents an energy storage device
"""
    energy_source: Optional[EnergySource]
    _height: float
    _width: float
class Inserter(StaticEntity, Electric):
"""
Represents an inserter that moves items between entities.
    Requires electricity to power
"""
    pickup_position: Optional[Position]
    drop_position: Position
    _width: float
    _height: float
class Filtered(BaseModel):
    filter: Optional[Any]
class UndergroundBelt(TransportBelt):
"""
An underground section of transport belt.
"""
    is_input: bool
    connected_to: Optional[int]
    _height: float
    _width: float
class MiningDrill(StaticEntity):
"""
Base class for mining drills that extract resources.
    The direction of the drill is where the drop_position is oriented towards
"""
    drop_position: Position
    resources: List[Ingredient]
class ElectricMiningDrill(MiningDrill, Electric):
"""
An electrically-powered mining drill.
"""
    _height: float
    _width: float
class BurnerInserter(Inserter, BurnerType):
"""
An inserter powered by burnable fuel.
"""
    _height: float
    _width: float
class BurnerMiningDrill(MiningDrill, BurnerType):
"""
A mining drill powered by burnable fuel.
"""
    _width = 2
    _height = 2
class Ammo(BaseModel):
    name: str
    magazine_size: Optional[int]
    reload_time: Optional[float]
class GunTurret(StaticEntity):
    turret_ammo: Inventory
    _height: float
    _width: float
    kills: Optional[int]
class AssemblingMachine(StaticEntity, Electric):
"""
A machine that crafts items from ingredients.
    Requires power to operate
"""
    recipe: Optional[Recipe]
    assembling_machine_input: Inventory
    assembling_machine_output: Inventory
    assembling_machine_modules: Inventory
    _height: float
    _width: float
class FluidHandler(StaticEntity):
"""
Base class for entities that handle fluids
"""
    connection_points: List[Position]
    fluid_box: Optional[Union[dict, list]]
    fluid_systems: Optional[Union[dict, list]]
class AdvancedAssemblingMachine(FluidHandler, AssemblingMachine):
"""
A second and third tier assembling machine that can handle fluids.
    Requires power to operate
    A recipe first needs to be set and then the input fluid source can be connected with pipes
"""
    _height: float
    _width: float
class MultiFluidHandler(StaticEntity):
"""
Base class for entities that handle multiple fluid types.
"""
    input_fluids: List[str]
    output_fluids: List[str]
    input_connection_points: List[IndexedPosition]
    output_connection_points: List[IndexedPosition]
    fluid_box: Optional[Union[dict, list]]
    fluid_systems: Optional[Union[dict, list]]
class FilterInserter(Inserter, Filtered):
"""
A inserter that only moves specific items
"""
    _height: float
    _width: float
class ChemicalPlant(MultiFluidHandler, AssemblingMachine):
"""
Represents a chemical plant that processes fluid recipes.
    Requires powering and accepts input fluids (from storage tanks etc) and solids (with inserters)
    Outputs either:
        solids (battery, plastic) that need to be extracted with inserters
        fluids (sulfuric acid, oil) that need to be extracted with pipes
    IMPORTANT: First a recipe needs to be set and then the fluid sources can be connected to the plant
"""
    _height: float
    _width: float
class OilRefinery(MultiFluidHandler, AssemblingMachine):
"""
An oil refinery for processing crude oil into products.
    Requires powering and accepts input fluids (from pumpjacks, storage tanks etc) and solids
    First a recipe needs to be set and then the fluid sources can be connected to the refinery
"""
    _height: float
    _width: float
class PumpJack(MiningDrill, FluidHandler, Electric):
"""
A pump jack for extracting crude oil. Requires electricity
    This needs to be placed on crude oil and oil needs to be extracted with pipes
    Oil can be sent to a storage tank, oil refinery or a chemical plant
    Oil can also be sent to assmbling machine to be made into oil barrels
    Important: The PumpJack needs to be placed on exact crude oil tiles
    
"""
    _height: float
    _width: float
class SolarPanel(ElectricalProducer):
"""
A solar panel for generating power from sunlight.
    This entity generated power during the day
    Thus it can be directly connected to a entity to power it
"""
    _height: float
    _width: float
class Boiler(FluidHandler, BurnerType):
"""
A boiler that heats water into steam.
"""
    steam_output_point: Optional[Position]
    _height: float
    _width: float
class HeatExchanger(Boiler):
"""
A nuclear heat exchanger that converts water to steam.
"""
class Generator(FluidHandler, StaticEntity):
"""
A steam generator that produces electricity.
"""
    _height: float
    _width: float
class Pump(FluidHandler, Electric):
"""
An electrically-powered fluid pump.
"""
    _height: float
    _width: float
class OffshorePump(FluidHandler):
"""
A pump that extracts water from water tiles. 
    Can be used in power generation setups and to supply water to chemical plants and oil refineries.
"""
    _height: float
    _width: float
class ElectricityPole(Entity, Electric):
"""
A power pole for electricity distribution.
"""
    flow_rate: float
    _height: float
    _width: float
    def __hash__(self):
class Furnace(Entity, BurnerType):
"""
A furnace for smelting items
"""
    furnace_source: Inventory
    furnace_result: Inventory
    _height: float
    _width: float
class ElectricFurnace(Entity, Electric):
"""
An electrically-powered furnace.
"""
    furnace_source: Inventory
    furnace_result: Inventory
    _height: float
    _width: float
class Chest(Entity):
"""
A storage chest.
"""
    inventory: Inventory
    _height: float
    _width: float
class StorageTank(FluidHandler):
"""
A tank for storing fluids.
    Can be used for inputs and outputs of chemical plants and refineries.
    Also can store water from offshore pumps.
"""
    _height: float
    _width: float
class RocketSilo(StaticEntity, Electric):
"""
A rocket silo that can build and launch rockets.
"""
    rocket_parts: int
    rocket_inventory: Inventory
    rocket_progress: float
    launch_count: int
    _width: float
    _height: float
    def __repr__(self) -> str:
class Rocket(Entity):
"""
A rocket that can be launched from a silo.
"""
    payload: Optional[Inventory]
    launch_progress: float
    def __repr__(self) -> str:
class Lab(Entity, Electric):
"""
A research laboratory.
"""
    lab_input: Inventory
    lab_modules: Inventory
    research: Optional[Any]
    _height: float
    _width: float
    def __repr__(self) -> str:
class Pipe(Entity):
"""
A pipe for fluid transport
"""
    fluidbox_id: int
    flow_rate: float
    contents: float
    fluid: Optional[str]
    _height: float
    _width: float
class Reactor(StaticEntity):
"""
A nuclear reactor
"""
    _height: float
    _width: float
class EntityGroup(BaseModel):
    id: int
    status: EntityStatus
    position: Position
    name: str
class WallGroup(EntityGroup):
"""
A wall
"""
    name: str
    entities: List[Entity]
class BeltGroup(EntityGroup):
"""
A connected group of transport belts.
"""
    belts: List[TransportBelt]
    inputs: List[Entity]
    outputs: List[Entity]
    inventory: Inventory
    name: str
    def __repr__(self) -> str:
    def __str__(self):
class PipeGroup(EntityGroup):
"""
A connected group of pipes.
"""
    pipes: List[Pipe]
    name: str
    def __repr__(self) -> str:
    def __str__(self):
class ElectricityGroup(EntityGroup):
"""
Represents a connected power network.
"""
    name: str
    poles: List[ElectricityPole]
    def __repr__(self) -> str:
    def __hash__(self):
    def __str__(self):
```
```methods
can_place_entity(entity: Prototype, direction: Direction = <Direction.UP: 0>, position: Position = Position(x=0.0, y=0.0)) -> bool
"""
Tests to see if an entity can be placed at a given position
:param entity: Entity to place from inventory
:param direction: Cardinal direction to place entity
:param position: Position to place entity
:return: True if entity can be placed at position, else False
"""

craft_item(entity: Prototype, quantity: int = 1) -> int
"""
Craft an item from a Prototype if the ingredients exist in your inventory.
:param entity: Entity to craft
:param quantity: Quantity to craft
:return: Number of items crafted
"""

extract_item(entity: Prototype, source: Union[Position, Entity], quantity=5) -> int
"""
Extract an item from an entity\'s inventory at position (x, y) if it exists on the world.
:param entity: Entity prototype to extract, e.g Prototype.IronPlate
:param source: Entity or position to extract from
:param quantity: Quantity to extract
:example extract_item(Prototype.IronPlate, stone_furnace.position, 5)
:example extract_item(Prototype.CopperWire, stone_furnace, 5)
:return The number of items extracted.
"""

get_connection_amount(source: Union[Position, Entity, EntityGroup], target: Union[Position, Entity, EntityGroup], connection_type: Prototype = <Prototype.Pipe: (\'pipe\', <class \'Pipe\'>)>) -> int
"""
Calculate the number of connecting entities needed to connect two entities, positions or groups.
:param source: First entity or position
:param target: Second entity or position
:param connection_type: a Pipe, TransportBelt or ElectricPole
:return: A integer representing how many entities are required to connect the source and target entities
"""

get_entities(entities: Union[Set[Prototype], Prototype] = set(), position: Position = None, radius: float = 1000) -> List[Entity]
"""
Get entities within a radius of a given position.
:param entities: Set of entity prototypes to filter by. If empty, all entities are returned.
:param position: Position to search around. Can be a Position object or "player" for player\'s position.
:param radius: Radius to search within.
:return: Found entities
"""

get_entity(entity: Prototype, position: Position) -> Entity
"""
Retrieve a given entity object at position (x, y) if it exists on the world.
:param entity: Entity prototype to get, e.g Prototype.StoneFurnace
:param position: Position where to look
:return: Entity object
"""

get_prototype_recipe(prototype: Union[Prototype, RecipeName, str]) -> Recipe
"""
Get the recipe (cost to make) of the given entity prototype.
:param prototype: Prototype to get recipe from
:return: Recipe of the given prototype
"""

get_research_progress(technology: Optional[Technology] = None) -> List[Ingredient]
"""
Get the progress of research for a specific technology or the current research.
:param technology: Optional technology to check. If None, checks current research.
:return The remaining ingredients to complete the research
"""

get_resource_patch(resource: Resource, position: Position, radius: int = 10) -> Optional[ResourcePatch]
"""
Get the resource patch at position (x, y) if it exists in the radius.
if radius is set to 0, it will only check the exact position for this resource patch.
:param resource: Resource to get, e.g Resource.Coal
:param position: Position to get resource patch
:param radius: Radius to search for resource patch
:example coal_patch_at_origin = get_resource_patch(Resource.Coal, Position(x=0, y=0))
:return: ResourcePatch if found, else None
"""

harvest_resource(position: Position, quantity=1, radius=10) -> int
"""
Harvest a resource at position (x, y) if it exists on the world.
:param position: Position to harvest resource
:param quantity: Quantity to harvest
:example harvest_resource(nearest(Resource.Coal), 5)
:example harvest_resource(nearest(Resource.Stone), 5)
:return: The quantity of the resource harvested
"""

insert_item(entity: Prototype, target: Union[Entity, EntityGroup], quantity=5) -> Entity
"""
Insert an item into a target entity\'s inventory
:param entity: Type to insert from inventory
:param target: Entity to insert into
:param quantity: Quantity to insert
:return: The target entity inserted into
"""

inspect_inventory(entity=None) -> Inventory
"""
Inspects the inventory of the given entity. If no entity is given, inspect your own inventory.
:param entity: Entity to inspect
:return: Inventory of the given entity
"""

launch_rocket(silo: Union[Position, RocketSilo]) -> RocketSilo
"""
Launch a rocket.
:param silo: Rocket silo
:return: Your final position
"""

move_to(position: Position, laying: Prototype = None, leading: Prototype = None) -> Position
"""
Move to a position.
:param position: Position to move to.
:return: Your final position
"""

nearest(type: Union[Prototype, Resource]) -> Position
"""
Find the nearest entity or resource to your position.
:param type: Entity or resource type to find
:return: Position of nearest entity or resource
"""

nearest_buildable(entity: Prototype, building_box: BuildingBox, center_position: Position, **kwargs) -> BoundingBox
"""
Find the nearest buildable area for an entity.

:param entity: Prototype of the entity to build.
:param building_box: The building box denoting the area of location that must be placeable.
:param center_position: The position to find the nearest area where building box fits
:return: BoundingBox of the nearest buildable area or None if no such area exists.
"""

pickup_entity(entity: Union[Entity, Prototype, EntityGroup], position: Optional[Position] = None) -> bool
"""
Pick up an entity if it exists on the world at a given position.
:param entity: Entity prototype to pickup, e.g Prototype.IronPlate
:param position: Position to pickup entity
:return: True if the entity was picked up successfully, False otherwise.
"""

place_entity(entity: Prototype, direction: Direction = <Direction.UP: 0>, position: Position = Position(x=0.0, y=0.0), exact: bool = True) -> Entity
"""
Places an entity e at local position (x, y) if you have it in inventory.
:param entity: Entity to place
:param direction: Cardinal direction to place
:param position: Position to place entity
:param exact: If True, place entity at exact position, else place entity at nearest possible position
:return: Entity object
"""

place_entity_next_to(entity: Prototype, reference_position: Position = Position(x=0.0, y=0.0), direction: Direction = <Direction.RIGHT: 2>, spacing: int = 0) -> Entity
"""
Places an entity next to an existing entity, with an optional space in-between (0 space means adjacent).
In order to place something with a gap, you must increase the spacing parameter.
:param entity: Entity to place
:param reference_position: Position of existing entity or position to place entity next to
:param direction: Direction to place entity from reference_position
:param spacing: Space between entity and reference_position
:example: place_entity_next_to(Prototype.WoodenChest, Position(x=0, y=0), direction=Direction.UP, spacing=1)
:return: Entity placed
"""

print(*args)
"""
Adds a string to stdout
:param args:
:return:
"""

rotate_entity(entity: Entity, direction: Direction = <Direction.UP: 0>) -> Entity
"""
Rotate an entity to a specified direction
:param entity: Entity to rotate
:param direction: Direction to rotate
:example rotate_entity(iron_chest, Direction.UP)
:return: Returns the rotated entity
"""

set_entity_recipe(entity: Entity, prototype: Union[Prototype, RecipeName]) -> Entity
"""
Sets the recipe of an given entity.
:param entity: Entity to set recipe
:param prototype: The prototype to create, or a recipe name for more complex processes
:return: Entity that had its recipe set
"""

set_research(*args, **kwargs)
"""
Call self as a function.
"""

set_research(technology: Technology) -> List[Ingredient]
"""
Set the current research technology for the player\'s force.
:param technology: Technology to research
:return: Required ingredients to research the technology.
"""

shift_entity(entity: Entity, direction: Union[Direction, Direction], distance: int = 1) -> Entity
"""
Calculate the number of connecting entities needed to connect two entities, positions or groups.
:param source: First entity or position
:param target: Second entity or position
:param connection_type: a Pipe, TransportBelt or ElectricPole
:return: A integer representing how many entities are required to connect the source and target entities
"""

sleep(seconds: int) -> bool
"""
Sleep for up to 15 seconds before continuing. Useful for waiting for actions to complete.
:param seconds: Number of seconds to sleep.
:return: True if sleep was successful.
"""
```
\end{lstlisting}

\subsection*{Example Memory}
Below is an example of 2 step memory that is input to the agent. The past policies $P$ and $O$ are formatted like user-assistant messages for chat models, where policies are assistant and observation is user messages. To improve the informativeness of logs, we mask out the overall game setup prints of past policies except for the last step. This makes sure the agent doesnt rely on past game states to make decisions

\begin{lstlisting}[language={}]
Assstant message
```python
water_pos = nearest(Resource.Water)
print(f"Found water source at {water_pos}")

# Place offshore pump
move_to(water_pos)
offshore_pump = place_entity(Prototype.OffshorePump, position=water_pos)
print(f"Placed offshore pump at {offshore_pump.position}")

boiler = place_entity_next_to(Prototype.Boiler, reference_position=offshore_pump.position, spacing = 4, direction = Direction.RIGHT)
print(f"Placed boiler at {boiler.position}")
steam_engine = place_entity_next_to(Prototype.SteamEngine, reference_position=boiler.position, spacing = 5, direction = Direction.RIGHT)
print(f"Placed steam engine at {steam_engine.position}")
# Connect offshore pump to boiler with pipes
pump_to_boiler = connect_entities(offshore_pump.position, boiler.position, Prototype.Pipe)
print(f"Connected offshore pump to boiler with pipes: {pump_to_boiler}")

# Connect boiler to steam engine with pipes
boiler_to_engine = connect_entities(boiler.position, steam_engine.position, Prototype.Pipe)
print(f"Connected boiler to steam engine with pipes: {boiler_to_engine}")
print(f"Current inventory {inspect_inventory()}")
print(f"Updated entities on the map: {get_entities()}")
```

User message
27015: ('Found water source at x=-12.0 y=0.0',)
27015: ('Placed offshore pump at x=-10.5 y=-0.5',)
27015: ('Placed boiler at x=-5.0 y=-0.5',)
27015: ('Placed steam engine at x=3.5 y=-0.5',)
27015: ('Connected offshore pump to boiler with pipes: [
    PipeGroup(fluid_system=1, position=x=-6.5 y=-2.5, status=EntityStatus.WORKING, pipes=[7 pipes], fluid=water)]',)
27015: ('Connected boiler to steam engine with pipes: [ PipeGroup(fluid_system=3, position=x=-2.5 y=-0.5, status=EntityStatus.EMPTY, pipes=[5 pipes])]',)
27015: ("Current inventory {'wooden-chest': 10, 'transport-belt': 200, 'burner-inserter': 10, 'small-electric-pole': 200, 'pipe': 88, 'boiler': 3, 'steam-engine': 1, 'burner-mining-drill': 10, 'electric-mining-drill': 10, 'offshore-pump': 2, 'stone-furnace': 5, 'assembling-machine-1': 5, 'coal': 200, 'stone': 50, 'iron-ore': 62, 'iron-plate': 19, 'copper-plate': 52, 'iron-gear-wheel': 22, 'electronic-circuit': 99}",)
27015: ("Updated entities on the map: <STALE_ENTITY_DATA_OMITTED>)

Assstant message
```python
# move to iron position
iron_pos = nearest(Resource.IronOre)
move_to(iron_pos)
# place a lectric mining drill
drill = place_entity(Prototype.ElectricMiningDrill, position = iron_pos)
print(f"Put a drill to mine iron at {drill.position}")

# power the drill
steam_engine = get_entity(Prototype.SteamEngine, position = Position(x = 3.5, y=-0.5))
poles = connect_entities(steam_engine,drill,Prototype.SmallElectricPole)
print(f"Used poles {poles} to power drill at {drill.position}")

print(f"Current inventory {inspect_inventory()}")
print(f"Updated entities on the map: {get_entities()}")
```

User Message
27015: ('Put a drill to mine iron at x=-11.5 y=19.5',)
27015: ('Used poles [ElectricityGroup(id=18, poles=[(x=-1.5,y=4.5),(x=-3.5,y=6.5),(x=3.5,y=1.5),...,(x=-9.5,y=12.5),(x=-11.5,y=14.5),(x=-13.5,y=16.5)], voltage=0)] to power drill at x=-11.5 y=19.5',)
27015: ("Current inventory {'wooden-chest': 10, 'transport-belt': 200, 'burner-inserter': 10, 'small-electric-pole': 191, 'pipe': 88, 'boiler': 3, 'steam-engine': 1, 'burner-mining-drill': 10, 'electric-mining-drill': 9, 'offshore-pump': 2, 'stone-furnace': 5, 'assembling-machine-1': 5, 'coal': 200, 'stone': 50, 'iron-ore': 62, 'iron-plate': 19, 'copper-plate': 52, 'iron-gear-wheel': 22, 'electronic-circuit': 99}",)
27015: ("Updated entities on the map: [
OffshorePump(name='offshore-pump', position=Position(x=-10.5, y=-0.5), direction=Direction.RIGHT, energy=0.0, tile_dimensions=TileDimensions(tile_width=1.0, tile_height=1.0), status=EntityStatus.WORKING, connection_points=[Position(x=-9.5, y=-0.5)], fluid_box=[{'name': 'water', 'amount': 100, 'temperature': 15}], fluid_systems=[49]), 
Boiler(fuel={}, name='boiler', position=Position(x=-5.0, y=-0.5), direction=Direction.RIGHT, energy=0.0, tile_dimensions=TileDimensions(tile_width=3.0, tile_height=2.0), warnings=['out of fuel'], status=EntityStatus.NO_FUEL, connection_points=[Position(x=-5.5, y=-2.5), Position(x=-5.5, y=1.5)], fluid_box=[{'name': 'water', 'amount': 200, 'temperature': 15}], fluid_systems=[49], steam_output_point=Position(x=-3.0, y=-0.5)), 
Generator(electrical_id=18, name='steam-engine', position=Position(x=3.5, y=-0.5), direction=Direction.RIGHT, energy=0.0, tile_dimensions=TileDimensions(tile_width=3.0, tile_height=5.0), warnings=['not receiving electricity', 'no input liquid', 'No fluid present in connections'], status=EntityStatus.NOT_CONNECTED, connection_points=[Position(x=6.0, y=-0.5), Position(x=1.0, y=-0.5)], fluid_box=[], fluid_systems=[]),
ElectricMiningDrill(electrical_id=18, name='electric-mining-drill', position=Position(x=-11.5, y=19.5), direction=Direction.UP, energy=0.0, tile_dimensions=TileDimensions(tile_width=3.0, tile_height=3.0), warnings=['not receiving electricity'], status=EntityStatus.NO_POWER, drop_position=Position(x=-11.5, y=17.5)),
PipeGroup(fluid_system=49, position=x=-6.5 y=-2.5, status=EntityStatus.FULL_OUTPUT, pipes=[7 pipes], fluid=water), 
PipeGroup(fluid_system=51, position=x=-2.5 y=-0.5, status=EntityStatus.EMPTY, pipes=[5 pipes]), 
ElectricityGroup(id=18, poles=[(x=-1.5,y=4.5),(x=-3.5,y=6.5),(x=3.5,y=1.5),...,(x=-9.5,y=12.5),(x=-11.5,y=14.5),(x=-13.5,y=16.5)], voltage=0)]",)

\end{lstlisting}


\end{document}